\documentclass[prd,twocolumn,aps, floatfix,preprintnumbers,nofootinbib, showpacs]{revtex4}

\usepackage{amsthm}   
\usepackage{amsmath}  
\usepackage{amssymb} 
\usepackage{mathrsfs}
\usepackage[all]{xy}
\usepackage[pdftex]{graphicx}
\usepackage{indentfirst}
\usepackage[pdftex]{hyperref}
\usepackage{verbatim}
\usepackage{float}
\usepackage{subfig}
\usepackage[justification=raggedright,singlelinecheck=false]{caption}

\begin{document}

\title{Quantization of Perturbations in an Inflating Elastic Solid}

\author{Michael Sitwell}
\affiliation{Department of Physics and Astronomy, University of British Columbia, Vancouver, BC, V6T 1Z1, Canada}

\author{Kris Sigurdson}
\affiliation{Department of Physics and Astronomy, University of British Columbia, Vancouver, BC, V6T 1Z1, Canada}

\date{\today}

\begin{abstract}

A sufficiently rigid relativistic elastic solid can be stable for negative pressure values and thus is capable of driving a stage of accelerated expansion. If a relativistic elastic solid drove an inflationary stage in the early Universe, quantum mechanically excited perturbations would arise in the medium. We quantize the linear scalar and tensor perturbations and investigate the observational consequences of having such an inflationary period. We find that slowly varying sound speeds of the perturbations and a slowing varying equation of state of the solid can produce a slightly red-tilted scalar power spectrum that agrees with current observational data. Even in the absence of non-adiabatic pressures, perturbations evolve on superhorizon scales, due to the shear stresses within the solid. As such, the spectra of perturbations are in general sensitive to the details of the end of inflation and we characterize this dependence.  Interestingly, we uncover here accelerating solutions for elastic solids with $1 + P/\rho$ significantly greater than $0$ that nevertheless have nearly scale-invariant scalar and tensor spectra. Beyond theoretical interest, this may allow for the possibility of viable inflationary phenomenology relatively far from the de Sitter regime.

\end{abstract}


\maketitle

\section{Introduction}

The inflationary paradigm, the existence of a brief period of accelerated expansion in the early Universe, provides an explanation for the observed homogeneity, isotropy and flatness of the Universe. On large scales it successfully accounts for the distribution of fluctuations seen in the cosmic microwave background (CMB) and the large-scale structure of the Universe. Inflation is often modelled  in terms of a scalar field slowly evolving in its potential.  Yet, the physical model of inflation is not  known and even within the context of scalar fields many  models are compatible with current observations.   It is worthwhile exploring whether or not other physical frameworks, more general than a scalar field,  can successfully account for a period of inflation.
 
In this paper, we build a model of inflation that describes the substance that drives inflation by a continuous medium that can be characterized by its macroscopic properties. The simplest model of a continuous medium in general relativity is a perfect fluid. To drive an accelerating expansion, the medium must have an equation of state $w\equiv P/\rho < -1/3$, where $\rho$ and $P$ are the energy density and pressure of the fluid, respectively. However, a perfect fluid with constant $w$  has a sound speed for longitudinal (density) waves of $c_s=\sqrt{w}$, so demanding that the fluid  drive an accelerated expansion formally results in an imaginary sound speed and an instability to small perturbations.

One generalization of a perfect fluid is a relativistic elastic solid. Elastic solids have a rigidity, and so can support both longitudinal and transverse waves. An elastic solid (both relativistic and nonrelativistic) can be characterized by a bulk modulus $\kappa$ that depends on the equation of state $w$, and shear modulus $\mu$ that determines how rigid the solid is. As in the nonrelativistic case, the longitudinal sound speed $c_s$ depends upon both $\kappa$ and $\mu$, while the transverse sound speed $c_v$ only depends upon $\mu$. In the relativistic case, a sufficiently rigid elastic solid can result in a real longitudinal sound speed $c_s$, even in cases where $w$ is negative enough to drive acceleration.

In this paper we describe a model of a homogeneous and isotropic elastic solid coupled to general relativity.  This model has previously been considered as a potential model of dark energy \cite{MossPert, Bucher} and recently similarities between a relativistic elastic solid and massive gravity have been noted \cite{massiveGravity}. In this work, we discuss in detail how an elastic solid can drive an inflationary epoch in the early Universe.\footnote{
The notion that a relativistic elastic solid could drive  inflation was first discussed in Ref.~\cite{Gruz} and more recently in Ref.~\cite{solidInflation} where an effective theory describing an elastic solid was developed.  Our work uses a different approach and treats the problem starting directly from the quadratic action for an elastic solid and includes an extended treatment of superhorizon evolution and reheating.  Reference~\cite{anisotropicSolid} has also recently examined the implications of anisotropic superhorizon evolution in an inflating elastic solid.}
 Linear perturbations in an elastic solid satisfy the equations of motion found in Ref.~\cite{MossPert}, which uses the framework for describing a macroscopic relativistic medium developed in Ref.~\cite{CarterFoundations}. We develop the quadratic action for a generic elastic medium, quantize the linear modes that are excited during inflation, and determine the spectra of  scalar and tensor modes produced by an inflationary stage driven by an elastic solid.

A novel feature of this model is that, in contrast to what typically occurs when the Universe is dominated by a single substance,  the anisotropic stress of the solid causes modes to evolve on superhorizon scales. As such, the final spectrum of superhorizon modes is sensitive to the manner in which inflation ends. We show here that the case where the sound speeds and equation of state are perfectly constant results in a blue-tilted scalar power spectrum, but if these quantities vary slowly in time then a red-tilted scalar power spectrum is possible.  While we do not specify a particular microphysical model for or formation mechanism of the elastic solid,  we note that relativistic elastic solids have been used to model a variety of physical systems  including networks of topological defects  \cite{rigidity,domainwalls}.   Interestingly, we find here that in models with slowly evolving material properties a scalar spectral index near $n_s \lesssim 1$, compatible with current observational constraints, can be found for $w$ relatively far from the nominal inflationary value $w \simeq -1$.

This paper is organized as follows: In Sections~\ref{einstSection} and \ref{elasticSection} we review the relevant Einstein equations and linearized perturbation equations for a relativistic elastic solid. In Section~\ref{actionSection} we derive the action for the scalar and tensor linear perturbations of a relativistic elastic solid. The superhorizon evolution in this model is discussed in Section~\ref{super} and its application to a period of inflation in the early Universe is discussed in Sections~\ref{inflationSection} to \ref{reheating}.

\section{Einstein Equations}
\label{einstSection}

We consider the flat Friedmann-Robertson-Walker (FRW) metric \footnote{We use signature (+,-,-,-) and units in which $c=\hbar=k_B=1$.}
\begin{equation}
		ds^2 = a^2(\eta)(\eta_{\alpha\beta} - h_{\alpha\beta}) dx^{\alpha}dx^{\beta}
\label{basicmetric}
\end{equation}
where $\eta$ is the conformal time, $\eta_{\alpha\beta}$ is the metric for Minkowski space and the tensor $h_{\alpha\beta}$ represents small perturbations to $\eta_{\alpha\beta}$. The energy density $\rho$ and pressure $P$ of the background in a flat universe can be expressed as
\begin{equation}
	\rho = \frac{\mathcal{H}^2}{l^2 a^2} \quad\quad P = -\frac{2\mathcal{H}' + \mathcal{H}^2}{3l^2 a^2}
\label{backgroundenergy}
\end{equation}
where a prime $'$ represents a derivative with respect to the conformal time $\eta$, $\mathcal{H}=a'/a=aH$, $H$ is the Hubble parameter, and $l=\sqrt{8\pi G/3}$ is the Planck length. We assume that we can parametrize the pressure by $P=w\rho$, where $w$ is referred to as the equation of state, so that with Eq.~(\ref{backgroundenergy}) we can form the differential equation for $\mathcal{H}$
\begin{equation}
\mathcal{H}' = -\frac{1+3w}{2}\mathcal{H}^2
\label{HDE}
\end{equation}
Using the Friedmann equation
\begin{equation}
	\rho' = -3\mathcal{H}\rho(1+w)
\label{drho}
\end{equation}
the relationship between $d P/d \rho$ and $w$ is found to be
\begin{equation}
	\frac{d P}{d \rho} = w - \frac{w'}{3\mathcal{H}(1+w)}
\end{equation}
We parametrize the metric in Eq.~(\ref{basicmetric}) as 
\begin{multline}
	ds^2 = a^2(\eta)\left[ (1+2\phi)d\eta^2 -2B,_i dx^i d\eta \right. \\
	\left. -( (1+ h/3)\delta_{ij} + 2E_{ij} ) dx^idx^j \right]
\label{metric}
\end{multline}
where $h$ is the trace of the spatial part of the metric perturbation and $E_{ij}$ is traceless. If we decompose the tensor $E_{ij}$ into scalar, vector, and tensor parts, then the scalar component of the spatial part of $h_{\alpha\beta}$ is
\begin{equation}
	h^S_{ij} = \frac{h}{3}\delta_{ij} + 2(\partial_i \partial_j - \frac{1}{3}\delta_{ij}\nabla^2)E = -2\psi\delta_{ij} + 2E,_{ij}
\label{hS}
\end{equation}
where $\psi\equiv -\frac{1}{6}h + \frac{1}{3}\nabla^2 E$ is the curvature perturbation and we denote the tensor part of $2E_{ij}$ by the conventional notation $h^T_{ij}$, which in addition to being traceless is transverse ($h^{Ti}_{j,i}=0$). 

The stress-energy tensor is parametrized in the standard form as
\begin{subequations}\label{grp}
	\begin{align}
	\delta T^0_0 &= \delta\rho \\
	\delta T^i_0 &= (\rho+P)v^i \\
	\delta T^i_j &= -(\delta P \delta^i_j + P \Pi^i_j)
\end{align}
\label{Tparam}
\end{subequations}
\!\!\!where $\delta\rho$ and $\delta P$ are the energy density and pressure perturbations, respectively, $v^i$ is the velocity perturbation, and $\Pi^i_j$ is the anisotropic stress.

The gauge-invariant Einstein equations for the scalar perturbations are
\begin{subequations}\label{grp}
\begin{equation}
	\nabla^2\Psi-3\mathcal{H}(\Psi'+\mathcal{H}\Phi) = \frac{3}{2}l^2 a^2 \delta\rho^{(gi)} \label{einst00}
\end{equation}
\begin{equation}
	\Psi'+\mathcal{H}\Phi = \beta v^{(gi)}
\label{einst0i}
\end{equation}
\begin{multline}	\Psi''+\mathcal{H}(\Phi'+2\Psi')+(\mathcal{H}^2+2\mathcal{H}')\Phi \\
+\frac{1}{3}\nabla^2(\Phi-\Psi) = \frac{3}{2}l^2 a^2 \delta P^{(gi)} \label{einstii}
\end{multline}
\begin{equation}
	\Psi - \Phi = 3 l^2 a^2 P \Pi
\label{einstij}
\end{equation}
\end{subequations}
where $\beta = \mathcal{H}^2-\mathcal{H}' = \frac{3}{2}l^2 a^2 (\rho+P)$, and the energy-momentum conservation equations are
\begin{subequations}\label{grp}
\begin{equation}
	\delta^{(gi)\prime} -(1+w)(\nabla^2 v^{(gi)}+3\Psi') +3\mathcal{H}\left(\frac{\delta P^{(gi)}}{\rho}-w\delta^{(gi)}\right) = 0
\label{energycons}
\end{equation}
\begin{multline}
	v^{(gi)\prime} + \mathcal{H}(1-3w)v^{(gi)} + \frac{w'}{1+w}v^{(gi)} \\ -\frac{\delta P^{(gi)}}{\rho+P} -\Phi - \frac{2}{3}\frac{w}{1+w}\nabla^2\Pi = 0
\label{velcons}
\end{multline}
\end{subequations}
where the gauge-invariant perturbation variables are defined as
\begin{subequations}
	\begin{equation}
		\Phi = \phi + \mathcal{H}(B-E') + (B-E')'
	\end{equation}
	\begin{equation}
		\Psi = \psi - \mathcal{H}(B-E')
	\end{equation}
	\begin{equation}
		\delta\rho^{(gi)} = \delta\rho + \rho'(B-E')
	\end{equation}
	\begin{equation}
		\delta P^{(gi)} = \delta P + P'(B-E')
	\end{equation}
	\begin{equation}
		v^{(gi)} = v + B-E'
	\end{equation}
\end{subequations}
noting that $\Pi$ is already a gauge-invariant quantity.

The sole Einstein equation for the tensor perturbations is 
\begin{equation}
(h^T)^{i\prime\prime}_j + 2\mathcal{H}(h^T)^{i\prime}_j - \nabla^2 (h^T)^i_j = 6 l^2 a^2 P (\Pi^T)^i_j
\label{heinst}
\end{equation}

\section{Elastic Solid}
\label{elasticSection}

In this section, we briefly summarize the formalism in Ref.~\cite{CarterFoundations} for a continuous relativistic medium and the findings of Ref.~\cite{MossPert} for the linear perturbations in an isotropic relativistic elastic solid.

As shown in Ref.~\cite{CarterFoundations}, the behaviour of a continuous relativistic medium can be described by the use of two different manifolds: a three-dimensional manifold $\mathcal{F}$ used to characterize the internal state of the medium and a four-dimensional spacetime manifold $\mathcal{M}$ used to describe its relativistic evolution. A projection $\mathcal{P}: \mathcal{M}\rightarrow\mathcal{F}$ is used to project timelike lines in $\mathcal{M}$ onto points in the material space on $\mathcal{F}$. This can be interpreted as projecting the worldline of a `particle' of the medium onto a single point in the material space. The internal properties of the medium are characterized through tensors defined on $\mathcal{F}$ that are then mapped onto $\mathcal{M}$ via the inverse image $\mathcal{P}^{-1}$. We use uppercase Latin letters $A,B,\ldots$ and lowercase Greek letters $\mu, \nu,\ldots$ to label the indices of tensors defined on $\mathcal{F}$ and $\mathcal{M}$, respectively. 

As the four-demensional projection tensor $\gamma^{\mu\nu}= g^{\mu\nu} - u^{\mu}u^{\nu}$ can be used to find the distance between adjacent particles in their local rest frame, $\gamma^{\mu\nu}$ and its material space counterpart $\gamma^{AB}$ characterize the strain of the medium. Working in material space, we assume that the energy density $\rho$ and pressure tensor $P^{AB}$ can be expressed in terms of the strain tensor $\gamma^{AB}$ and are related by
\begin{subequations}
	\begin{equation}
		\partial(\sqrt{|\gamma|} \rho) = -\frac{\sqrt{|\gamma|}}{2} P^{AB} \partial\gamma_{AB}
	\end{equation}
	\begin{equation}
		\partial(\sqrt{|\gamma|} P^{AB}) = -\frac{\sqrt{|\gamma|}}{2} E^{ABCD} \partial\gamma_{CD}
	\label{stresseqn}
	\end{equation}
\end{subequations}
in close analogy to the classical case, where $|\gamma|$ is the determinant of $\gamma_{AB}$. The elasticity tensor $E^{ABCD}$ has been introduced in Eq.~(\ref{stresseqn}) to relate stress and strain tensors, as in the classical case.

As shown in Ref.~\cite{MossPert}, specifying the pressure and elasticity tensors is sufficient for describing the behaviour of linear perturbations in a relativistic elastic solid. By relating derivatives in $\mathcal{F}$ and $\mathcal{M}$, the spacetime pressure and elasticity tensors for an isotropic elastic solid are
\begin{subequations}
	\begin{equation}
		P^{\mu\nu} = P \gamma^{\mu\nu}
	\label{pressureElastic}
	\end{equation}
	\begin{multline}
	E^{\mu\nu\rho\sigma} = \Sigma^{\mu\nu\rho\sigma} + \left((\rho+P)\frac{d P}{d \rho}-P\right)\gamma^{\mu\nu}\gamma^{\rho\sigma} \\+2P\gamma^{\mu(\rho}\gamma^{\sigma)\nu}
	\label{elasticTensor}
	\end{multline}
\end{subequations}
where $P$ is the pressure scalar and $\Sigma^{\mu\nu\rho\sigma}$ is the shear tensor given by
\begin{equation}
	\Sigma^{\mu\nu\rho\sigma} = 2\mu\left(\gamma^{\mu(\rho}\gamma^{\sigma)\nu}-\frac{1}{3}\gamma^{\mu\nu}\gamma^{\rho\sigma}\right)
\end{equation}
with $\mu$ being the shear modulus. For a perfectly elastic medium, the stress-energy tensor is related to the pressure tensor by
\begin{equation}
	T^{\mu\nu} = \rho u^{\mu}u^{\nu} + P^{\mu\nu}
\label{stressElastic}
\end{equation}
where $u^{\mu}$ are flow vectors tangent to worldlines.

An elastic solid has a resistance to compressive and shearing motions and thus can support both longitudinal and transverse waves, which travel at speeds $c_s$ and $c_v$, respectively. In both the relativistic and nonrelativistic cases, $c_s$ is dependent upon both the bulk modulus $\kappa$ and the shear modulus $\mu$, while $c_v$ is dependent only upon $\mu$. In the nonrelativistic case, the sound speeds and bulk modulus are given by \cite{Landau}
\begin{equation}
	c_s^2 = \frac{\kappa +\frac{4}{3}\mu}{\rho} \quad\quad c_v^2 = \frac{\mu}{\rho} \quad\quad \kappa = \rho\frac{d P}{d \rho}
\end{equation}
where the energy density $\rho$ is dominated by the mass contribution in the nonrelativistic limit. The sound speeds for the relativistic case can be found by making the substitution $\rho\rightarrow \rho+P$ so that \cite{CarterSpeed}
\begin{equation}
	c_s^2 = \frac{d P}{d \rho} + \frac{4}{3}c_v^2 \quad\quad\quad c_v^2 = \frac{\mu}{\rho + P}
\label{soundspeeds}
\end{equation}
and the bulk modulus is now given by $\kappa=(\rho+P)d P/d\rho$. We can see that even in the case where $d P/d\rho$ is negative, a real value for longitudinal sound speed $c_s$ can be obtained if the rigidity is sufficiently large.

We now examine the perturbations in the elastic solid. Perturbations in a continuous medium can be described by a shift vector $\xi^{\alpha}=\xi^{\alpha}(x^{\beta}_0)$, so that if a particle in a medium is at position $x^{\beta}_0$ when no perturbations are present, then the particle would be at position $x^{\alpha}(x^{\beta}_0)=\xi^{\alpha}(x^{\beta}_0)+x^{\alpha}_0$ when perturbations are present. By use of the above equations, it can be shown that the linear perturbations that arise in the stress-energy tensor can be written in terms of the shift vector $\xi^{\alpha}$ as \cite{MossPert}
\begin{subequations}\label{grp}
	\begin{equation}
		\delta T^0_0 = -(\rho+P)\left(\xi^k,_k + \frac{h}{2}\right)
	\end{equation}
	\begin{equation}
		\delta T^i_0 = (\rho+P)\xi^{i\prime}
	\end{equation}
	\begin{multline}
		\delta T^i_j = \frac{d P}{d \rho}(\rho+P)\left( \xi^k,_k + \frac{h}{2} \right)\delta^i_j \\
		+ \mu \left( 2\xi^{(i},_{j)}+ h^i_j -\frac{2}{3}\delta^i_j\left(\xi^k ,_k + \frac{h}{2} \right)\right)
	\end{multline}
\end{subequations}
By comparing these equations to the standard parametrizations given in Eq.~(\ref{Tparam}), we can make the following identifications
\begin{subequations}\label{grp}
	\begin{align}
		\delta\rho &= -(\rho+P)\left(\xi^k,_k + \frac{h}{2}\right) \\
		v^i &= \xi^{i\prime} \\
		\delta P &= \frac{d P}{d \rho} \delta\rho \label{deltaP} \\
		\Pi^i_j &= -\frac{\mu}{P} \left( 2\xi^{(i},_{j)}+ h^i_j -\frac{2}{3}\delta^i_j\left(\xi^k ,_k + \frac{h}{2} \right)\right)
	\end{align}
\label{elasticParams}
\end{subequations}
\!\!We note that Eq.~(\ref{deltaP}) implies that entropy perturbations are not present in the solid in the sense that the pressure perturbation is fully specified by the energy density and not the entropy. Taking the scalar parts of these equations yields
\begin{subequations}\label{grp}
	\begin{align}
		\delta & = -(1+w)(\nabla^2\xi^S-3\psi+\nabla^2E) \label{deltaEqn} \\
		v & = \xi^{S\prime} \\
		\nabla^2\Pi & = 2c_v^2(1+w^{-1})\left[-\nabla^2\xi^S-\nabla^2E\right]
\end{align}
\end{subequations}
where $\xi^S$ is the scalar part of the shift vector. Using Eq.~(\ref{deltaEqn}), we can rewrite the anisotropic stress as
\begin{align}
\nabla^2\Pi &= 2c_v^2(1+w^{-1}) \left[\frac{\delta}{1+w}-3\psi\right] \nonumber \\ &= -6c_v^2(1+w^{-1})\zeta
\label{pieqn2}
\end{align}
where we have identified the gauge-invariant variable $\zeta$ as
\begin{equation}
\zeta \equiv \psi + \mathcal{H} \frac{\delta\rho}{\rho'}
\label{zetaeqn}
\end{equation}
which can be interpreted as the curvature perturbation on uniform density hypersurfaces or as the density perturbation on uniform curvature hypersurfaces.

The tensor perturbations are simple in comparison. Equation~(\ref{elasticParams}) implies that the tensor part of the anisotropic stress is simply
\begin{equation}
	(\Pi^{T})^i_j = -\frac{\mu}{P} (h^T)^i_j
\label{Pitensor}
\end{equation}

Having characterized the general properties of our material, we can now begin to examine how perturbations are excited in an elastic solid.

\section{Action}
\label{actionSection}

To quantize the linear perturbations in the elastic solid, we start with its action and perturb it to second order in the perturbation variables to yield linear equations of motion. We decompose the action as $S = S_{\rm{m}} + S_{\rm{gr}}$, where $S_{\rm{m}}$ and $S_{\rm{gr}}$ are the matter and gravitational parts of the action, respectively. The gravitational part of the action is given by
\begin{equation}
	S_{\rm{gr}} = -\frac{1}{6l^2}\int \mathcal{R} \sqrt{-g} d^4x
\end{equation}
where $g\equiv \text{det}(g_{\mu\nu})$ and $\mathcal{R}$ is the Ricci scalar. The matter part of the action for a continuous medium is given by \cite{Fock}
\begin{equation}
	S_{\rm{m}} = -\int \rho^{\text{tot}} \sqrt{-g}d^4x
\label{fullaction}
\end{equation}
In the above equation, $\rho^{\text{tot}}$ is the total energy density, which we decompose as $\rho^{\text{tot}} = \rho^{\text{tot}}_f + \rho^{\text{tot}}_e$, where $\rho^{\text{tot}}_f$ is the energy density corresponding to a perfect fluid and $\rho^{\text{tot}}_e$ is the additional energy density arising from shear stresses in the elastic solid. The perfect fluid part of the action taken to second order in the perturbation variables can be expressed as \cite{Mukhanov}
\begin{multline}
	\delta_2 S_f = - \int \left[ \rho\frac{\delta_2\sqrt{-g}}{\sqrt{-g_0}} + (\rho+P)\left(\frac{\delta_1n}{n_0}\frac{\delta_1\sqrt{-g}}{\sqrt{-g_0}}+\frac{\delta_2n}{n_0}\right) \right. \\ \left. + \frac{1}{2}\frac{d P}{d \rho}(\rho+P)\left(\frac{\delta_1n}{n_0}\right)^2 \right] \sqrt{-g_0} d^4x
\end{multline}
where here the subscript $0$ indicates the background value, $\delta_1$ and $\delta_2$ denote the terms in a variable containing first and second order perturbations, respectively, and $n$ is the number density. For a relativistic isotropic elastic solid, $\rho^{\text{tot}}_e$ is given by \cite{Bucher}
\begin{equation}
	\rho^{\text{tot}}_e = \frac{P^2}{4\mu}\Pi_{ij}\Pi^{ij}
\label{elasticenergy}
\end{equation}
so that the action for the elastic part perturbed to second order is
\begin{equation}
	\delta_2 S_e = -\int \frac{a^4 P^2}{4\mu}\Pi_{ij}\Pi^{ij} d^4x
\label{elasticAction}
\end{equation}

\subsection{Quantization of Scalar Modes}
\label{scalarModes}

Using the expressions for the action as described above, the scalar part of the perfect fluid action, including the gravitational part, perturbed to second order can be found to be \cite{Mukhanov}
\begin{widetext}
\begin{multline}
	\delta_2 S_f + \delta_2 S_{\rm{gr}}  = \frac{1}{6l^2}  \int a^2 \left[ -6\left(\psi^{\prime 2}+2\mathcal{H}\phi\psi'+\left(\mathcal{H}^2-\frac{\beta}{3\frac{d P}{d\rho}}\right)\phi^2 \right) -4(\psi'+\mathcal{H}\phi)\nabla^2(B-E') \nonumber \right. \\ \left.  -2\psi,_i (2\phi,_i - \psi,_i) +2\beta(v^{,i}+B,_i)(v^{,i}+B,_i) - 2\beta \frac{d P}{d \rho} \left(3\psi-\nabla^2E-\nabla^2\xi^S + \frac{\phi}{\frac{d P}{d \rho}}\right)^2 \right] d^4x
\label{act1}
\end{multline}
\end{widetext}
We now wish to put the action in canonical form. We will work in the comoving gauge where $v=B=0$. The main reasons for this choice of gauge are that the action above is simplified greatly in this gauge and that the gauge-invariant variable $R$, defined by
\begin{equation}
	R \equiv \mathcal{H}v + \psi
\label{Reqn}
\end{equation}
in the comoving gauge is simply related to the metric perturbation $\psi$ by $R=\psi$ and so represents the curvature perturbation in this gauge. If we are able to form an expression solely in terms of $\psi$ in this gauge, the gauge-invariant expression can then be trivially found by substituting $R$ for $\psi$. In the comoving gauge, the Einstein equations (\ref{einst00}) and (\ref{einst0i}) and the momentum conservation equation (\ref{velcons}) are
\begin{subequations}\label{grp}
\begin{equation}
	\nabla^2(\psi + \mathcal{H}E') -3\mathcal{H}(\psi'+\mathcal{H}\phi) - \frac{3}{2}\mathcal{H}^2\delta = 0
\label{einst00co}
\end{equation}
\begin{equation}
	\psi'+\mathcal{H} \phi = 0
\label{einst0ico}
\end{equation}
\begin{equation}
	\frac{\frac{d P}{d \rho}}{1+w}\delta + \phi + \frac{2}{3}\frac{w}{1+w}\nabla^2\Pi = 0
\label{veqn2}
\end{equation}
\end{subequations}
Using Eqs.~(\ref{einst0ico}) and (\ref{veqn2}), the fluid part of the action in the comoving gauge becomes 
\begin{multline}
		\delta_2 S_f + \delta_2 S_{\rm{gr}}  = \frac{1}{3l^2}  \int \left[	 \frac{3(1+w)}{\frac{d P}{d\rho}} \psi'^2 - 3(1+w)\psi_{,i}^2 \right. \\ \left. - \frac{4}{3}\frac{w^2}{1+w}\frac{\mathcal{H}^2}{\frac{d P}{d\rho}}(\nabla^2\Pi)^2 \right] d^4x
\end{multline}
where a total derivative term has been dropped.

We now turn our attention to the additional part of the action for an elastic solid given in Eq.~(\ref{elasticAction}). For the scalar part of the anisotropic stress tensor, we have $(\Pi^S)_{ij}(\Pi^S)^{ij} = \frac{2}{3}(\nabla^2\Pi)^2 + \text{(total derivative term)}$, so the scalar part of the elastic part of the action is
\begin{equation}
\delta_2 S_e = -\frac{1}{6l^2} \int  \frac{w^2a^2\mathcal{H}^2}{c_v^2(1+w)} (\nabla^2\Pi)^2 d^4x
\end{equation}
where we have used the sounds speeds in Eq.~(\ref{soundspeeds}). We can then write the total action for the scalar perturbations as
\begin{multline}
	\delta_2 S = \frac{1}{6l^2}  \int a^2 \left[	 \frac{3(1+w)}{\frac{d P}{d \rho}}\psi'^2 - 3(1+w)\psi_{,i}^2 \right. \\ \left. -\frac{c_s^2w^2\mathcal{H}^2}{c_v^2(1+w)\frac{d P}{d \rho}} (\nabla^2\Pi)^2 \right] d^4x
\label{act3}
\end{multline}
We can express $\nabla^2\Pi$ as a function of $\psi$ by using Eqs.~(\ref{pieqn2}), (\ref{einst0ico}), and (\ref{veqn2}), which yields
\begin{equation}
	\nabla^2\Pi = 2\frac{c_v^2}{c_s^2}\frac{1+w}{w}\left[\frac{\psi'}{\mathcal{H}} - 3\frac{d P}{d\rho} \psi\right]
\label{Piscalar}
\end{equation}
Using this expression in the action above gives
\begin{multline}
		\delta_2 S = \frac{1}{3l^2}  \int z^2\left[R'^2 - c_s^2 R_{,i}^2 - 4c_v^2\beta R^2 \right. \\ \left. -4 c_v^2\mathcal{H} \left(\frac{(c_v^2)'}{c_v^2}-\frac{(c_s^2)'}{c_s^2}\right)R^2 \right] d^4x	
\label{Ractionmid}
\end{multline}
where we have cast the action into a gauge-invariant form and a total derivative term has been dropped. We have defined $z$ by 
\begin{equation}
	z \equiv \frac{a \sqrt{\beta}}{c_s \mathcal{H}} = \frac{a}{c_s}\sqrt{\frac{3}{2}(1+w)}
\label{zdef}
\end{equation}
We can now define the canonical variable $u$ as
\begin{equation}
	u \equiv \sqrt{\frac{2}{3l^2}} zR
\label{udef}
\end{equation}
so that the action becomes
\begin{equation}
	\delta_2 S = \frac{1}{2}  \int \left[ u'^2 -c_s^2 u_{,i}^2 -m^2_{\text{eff},S}(\eta)u^2 \right] d^4x
\label{uaction}
\end{equation}
where another total derivative term has been dropped and the effective mass is
\begin{equation}
	m^2_{\text{eff},S}(\eta) \equiv - \frac{z''}{z} + 4c_v^2\beta + 4c_v^2\mathcal{H} \left(\frac{(c_v^2)'}{c_v^2}-\frac{(c_s^2)'}{c_s^2}\right)
\label{meff}
\end{equation}
Varying the action with respect to $u$ leads to the equation of motion
\begin{equation}
	u'' - c_s^2\nabla^2u + m_{\text{eff},S}^2(\eta)u = 0
\label{ueom}
\end{equation}

It is clear that the action in Eq.~(\ref{uaction}) has the same form as the action for a harmonic oscillator with time-dependent mass, so we may use the same quantization procedure as is used to quantize a harmonic oscillator. The conjugate momentum $\pi$ to $u$ is
\begin{equation}
	\pi = \frac{\partial\mathcal{L}}{\partial u'} = u'
\end{equation}
We now promote $u$ and $\pi$ to operators $\hat{u}$ and $\hat{\pi}$ and impose the commutation relations
\begin{align}
	[\hat{u}(\textbf{x},\eta),\hat{u}(\tilde{\textbf{x}},\eta)] & = [\hat{\pi}(\textbf{x},\eta),\hat{\pi}(\tilde{\textbf{x}},\eta)] = 0 \nonumber \\
	[\hat{u}(\textbf{x},\eta),\hat{\pi}(\tilde{\textbf{x}},\eta)] & = i\delta(\textbf{x}-\tilde{\textbf{x}})
\label{comm}
\end{align}
Using the Fourier conventions
\begin{equation}
f(\textbf{x}) = \int \frac{d^3k}{(2\pi)^{3/2}} f_{\textbf{k}} e^{i\textbf{k}\cdot\textbf{x}}
\end{equation}
we can write $\hat{u}_{\textbf{k}}$ in terms of the creation and annihilation operators $\hat{a}_{\textbf{k}}^{\dagger}$ and $\hat{a}_{\textbf{k}}$ as $\hat{u}_{\textbf{k}}(\eta) =( \hat{a}_{\textbf{k}}\chi_{\textbf{k}}^*(\eta) + \hat{a}_{-\textbf{k}}^{\dagger}\chi_{\textbf{k}}(\eta)  )/\sqrt{2}$ so that
\begin{equation}
	\hat{u} (\textbf{x},\eta) = \frac{1}{\sqrt{2}} \int \frac{d^3 k}{(2\pi)^{3/2}}\left[ \hat{a}_{\textbf{k}} \chi_{\textbf{k}}^*(\eta) e^{i\textbf{k}\cdot\textbf{x}} + \hat{a}^{\dagger}_{\textbf{k}} \chi_{\textbf{k}}(\eta) e^{-i\textbf{k}\cdot\textbf{x}} \right]
\label{decomp}
\end{equation}
and the mode function $\chi_{\textbf{k}}(\eta)$ obeys
\begin{equation}
	\chi_{\textbf{k}}''+\left[c_s^2k^2+m_{\text{eff},S}^2\right]\chi_{\textbf{k}} = 0
\label{chieom}
\end{equation}
The commutation relations in Eq.~(\ref{comm}) imply the normalization
\begin{equation}
	\chi_{\textbf{k}}'\chi_{\textbf{k}}^* - \chi_{\textbf{k}} \chi_{\textbf{k}}^{*\prime} = 2i
\label{norm}
\end{equation}

Once we solve for the mode function $\chi_{\textbf{k}}$ from the differential equation in Eq.~(\ref{chieom}), subject to the the normalization condition above, we can calculate the power spectrum for the scalar perturbations $\mathcal{P}_{R}(k)=|R_k|^2 k^3/2\pi^2$. We will see in Section~\ref{super} that the perturbations associated with a single scalar mode with wavevector $\textbf{k}$ is anisotropic, due to the presence of anisotropic stress. However, each mode has the same evolution (the solution for $\chi_{\textbf{k}}$ in Eq.~(\ref{chieom}) is the same for each $\textbf{k}$ with the same magnitude $k$). As discussed in Section~\ref{super}, we will assume that expectation values are isotropic, so that $\langle u_{\textbf{k}}u_{\tilde{\textbf{k}}}\rangle = |u_k|^2\delta(\textbf{k}+\tilde{\textbf{k}})$. As a result, the integrand of the integral over $d^3k$ in the two-point correlation function for the operator $\hat{u}(\textbf{x}, \eta)$ with the vacuum state will depend only on $k$, so we can trivially integrate over the solid angle $d\Omega_k$, so that
\begin{equation}
\langle 0 | \hat{u}(\textbf{x},\eta)\hat{u}(\tilde{\textbf{x}},\eta) | 0 \rangle = \int dk \frac{k^2}{4\pi^2}|\chi_k(\eta)|^2 \frac{\text{sin}(k|\textbf{x}-\tilde{\textbf{x}}|)}{k|\textbf{x}-\tilde{\textbf{x}}|}
\end{equation}
We can now identify the power spectrum for $u$ as $\mathcal{P}_{u}(k) = \frac{k^3}{4\pi^2}|\chi_k|^2$ and using Eq.~(\ref{udef}) the power spectrum for $R$ will be
\begin{equation}
\mathcal{P}_{R}(k) = \frac{3l^2k^3}{8\pi^2z^2}|\chi_k|^2
\label{PReqn}
\end{equation}

\subsection{Quantization of Tensor Modes}

We now turn our attention to the tensor modes. Using the tensor part of the metric in Eq.~(\ref{metric}) to calculate the tensor part of the Ricci scalar $\mathcal{R}$, the gravitational part of the action can be found to be
\begin{equation}
	\delta_2 S_{\rm{gr}} = \frac{1}{24 l^2} \int a^2 \left[ (h^{T})^{i\prime}_j (h^T)^{j\prime}_i - (h^T)^i_{j,l}(h^T)^{j,l}_i \right] d^4x
\end{equation}
The only contribution to the matter part of the action is from the tensor part of the anisotropic stress, given by Eq.~(\ref{Pitensor}), which using the elastic part of the action in Eq.~(\ref{elasticAction}) is
\begin{equation}
		\delta_2 S_e = -\int \frac{a^4\mu}{4P} (h^T)^i_j (h^T)^j_i d^4x
\end{equation}
With the transverse sound speed given in Eq.~(\ref{soundspeeds}), the total action becomes
\begin{multline}
	\delta_2 S = \frac{1}{24 l^2} \int a^2 \left[ (h^{T})^{i\prime}_j (h^{T})^{j\prime}_i - (h^{T})^i_{j,l} (h^{T})^{j,l}_i \right. \\ \left. - 4c_v^2\beta (h^{T})^i_j (h^{T})^j_i \right] d^4x
\end{multline}
It is convenient to express $(h^{T})^i_j$ in terms of the individual polarization states $h^T_p$, where $(h^{T})^i_j (h^{T})^j_i = 2\sum_p (h_p^T)^2$, so that the action for each polarization state is
\begin{equation}
	\delta_2 S = \frac{1}{12 l^2} \int a^2 \left[ (h_p^T)'^2 - (h_p^T)_{,i}^2 - 4c_v^2\beta (h_p^T)^2 \right] d^4x
\end{equation}
We can define the canonical variable $U_p$ for the tensor perturbations as
\begin{equation}
U_p = \frac{a h_p^T}{\sqrt{6l^2}}
\end{equation}
so the action becomes
\begin{equation}
	\delta_2 S = \frac{1}{2} \int \left[ U_p'^2 - U_{p,i}^2 - m^{2}_{\text{eff},T}U_p^2 \right] d^4x
\label{tensoraction}
\end{equation}
where a total derivative has been dropped and the effective mass for the tensor modes is given by
\begin{equation}
	m^{2}_{\text{eff},T} = - \frac{a''}{a} + 4c_v^2\beta
\label{meffTensor}
\end{equation}
As with the scalar perturbations, the action for the tensor perturbations has the same form as the action of a harmonic oscillator with time-dependent mass. We can therefore use the same quantization procedure as was used with the scalar perturbations. We promote the canonical variable $U_p$ to an operator and write it in terms of creation and annihilation operators as
\begin{multline}
		\hat{U}_p (\textbf{x},\eta) = \frac{1}{\sqrt{2}} \int \frac{d^3 k}{(2\pi)^{3/2}} \bigg[ \hat{a}_{\textbf{k}} X_{\textbf{k}}^*(\eta) e^{i\textbf{k}\cdot\textbf{x}}  \\ + \hat{a}^{\dagger}_{\textbf{k}} X_{\textbf{k}}(\eta) e^{-i\textbf{k}\cdot\textbf{x}} \bigg]
\label{Xdecomp}
\end{multline}
and so the equation of motion for the mode function $X_{\textbf{k}}$ is
\begin{equation}
	X_{\textbf{k}}''+ \left[ k^2 + m^{2}_{\text{eff},T} \right] X_{\textbf{k}} = 0
\label{tensoreom}
\end{equation}
The commutation relations analogous to Eq.~(\ref{comm}) for the tensor case yields the normalization condition
\begin{equation}
	X_{\textbf{k}}'X_{\textbf{k}}^* - X_{\textbf{k}} X_{\textbf{k}}^{*\prime} = 2i
\label{normtensor}
\end{equation}

Accounting for both polarization states, the two-point function for the tensor perturbations with the vacuum state is then
\begin{multline}
\langle 0 | (\hat{h}^T)^i_j(\textbf{x},\eta) (\hat{h}^T)^j_i(\tilde{\textbf{x}},\eta) | 0 \rangle =  \int dk \frac{6l^2}{\pi^2 a^2} k^2|X_k(\eta)|^2 \\ \times \frac{\text{sin}(k|\textbf{x}-\tilde{\textbf{x}}|)}{k|\textbf{x}-\tilde{\textbf{x}}|}
\end{multline}
with which we can identify the tensor power spectrum $\mathcal{P}_T$ as 
\begin{equation}
\mathcal{P}_T(k) = \frac{6l^2 k^3}{\pi^2 a^2}|X_k|^2
\label{Ph}
\end{equation}

\section{Superhorizon Evolution}
\label{super}

An interesting phenomena in this model is that both $\zeta$ and $R$ evolve on superhorizon scales, even when the elastic solid is the only substance present in the Universe. Typically, this type of superhorizon evolution only arises in the presences of a nonadiabatic pressure $\delta P_{\text{nad}} = \delta P - (d P/d\rho)\delta\rho$. From Section~\ref{elasticSection}, we saw that $\delta P/\delta\rho = d P/d\rho$ for an elastic solid, so the nonadiabatic pressure vanishes. However, the addition of the anisotropic stress in the elastic solid adds another type of stress to the system, which causes superhorizon evolution in a similar manner to cases when nonadiabatic pressures are present.

In the standard case when only adiabatic and isotropic pressures are present, both $\zeta$ and $R$ remain approximately constant on superhorizon scales because once smoothed on a scale much larger than the horizon, each patch of the Universe smaller than the smoothing scale evolves approximately like a separate unperturbed FRW universe.  This idea is known as the `separate universe approach' \cite{separateUniverse}. The locally defined expansion $\tilde{\theta}(\textbf{x},t)$ with respect to coordinate time $t$ is given by
\begin{equation}
\tilde{\theta}(\textbf{x},t) = 3H - 3\dot{\psi}(\textbf{x},t) + \nabla^2\sigma(\textbf{x},t)
\label{localH}
\end{equation}
where an overdot denotes a derivative with respect to coordinate time and $\sigma=\dot{E}-B$ is the (local) shear. Considering a flat slicing ($\psi=0$), the local expansion will be equal to the background value if we can safely neglect the effects of the shear on large scales. In this case, since the (total) energy density evolves according to the local energy conservation equation, which to linear order is
\begin{equation}
\dot{\rho}^{\text{tot}}(\textbf{x},t) = - (\tilde{\theta}(\textbf{x},t) +\nabla^2 v(\textbf{x},t))(\rho^{\text{tot}}(\textbf{x},t) + P^{\text{tot}}(\textbf{x},t))
\end{equation}
then after smoothing on superhorizon scales, the local energy density at each location will (approximately) follow the same unperturbed FRW evolution. Therefore, the difference between the energy density perturbations at different locations will be kept approximately constant in time and since $\zeta$ is proportional to the energy density perturbation in a flat slicing, $\zeta$ will be approximately constant on superhorizon scales.\footnote{A similar argument can be made for $R$, which is proportional to $v$ in a spatially flat slicing, by using the local conservation of momentum.}

As was shown in Ref.~\cite{conservedCosmologicalPerturbations}, when the anisotropic stress is neglected, the shear is in fact negligible on large scales. However, the anisotropic stress acts as a source term for the shear (see Eq.~(31) of Ref.~\cite{cosmologicalModels}), causing the shear to be non-negligible on superhorizon scales in the case of an elastic solid. If the shear cannot be neglected, then different locations in the Universe after smoothing on superhorizon scales will not evolve as an unperturbed FRW universe owing to the fact that a FRW spacetime is shear free and in general the local expansion will be position dependent in a flat slicing.

Working in the gauge where $\psi=B=0$ so that the shear is $\sigma = \dot{E}$, the trace-free part of the spatial components  of the Einstein equations in Fourier space is
\begin{equation}
\dot{\sigma}_{\textbf{k}} + 3H\sigma_{\textbf{k}} - \frac{k^2}{a^2}\phi_{\textbf{k}} = 3wH^2 \Pi_{\textbf{k}}
\label{shearevolution}
\end{equation}
As we can see, the shear is indeed sourced by the anisotropic stress and will evolve on superhorizon scales unless the anisotropic stress is negligible on these scales. It is easily seen from Eq.~(\ref{pieqn2}) that for an elastic solid the anisotropic stress is significant (i.e. comparable to the energy density and pressure perturbations) on superhorizon scales since in a spatially flat slicing $\Pi_{\textbf{k}} = -2(c_v^2/w)\delta_{\textbf{k}}$. Therefore,  we will have $|\Pi_{\textbf{k}}|\sim |\delta_{\textbf{k}}|$ in this slicing for a sufficiently rigid solid and the shear will evolve as
\begin{equation}
\dot{\sigma}_{\textbf{k}} + 3H\sigma_{\textbf{k}} - \frac{k^2}{a^2}\phi_{\textbf{k}} = -6c_v^2H^2 \delta_{\textbf{k}}
\end{equation}
so that the shear is sourced by the (non-negligible) density perturbations on superhorizon scales when viewed in this gauge.

If we consider a mode with wavevector $\textbf{k}=(0,0,k)$ then the scalar part of the anisotropic stress tensor in Fourier space, given by $(\Pi_{\textbf{k}}^S)_{ij}=(-k_i k_j/k^2 + \delta_{ij}/3)\Pi_{\textbf{k}}$, is $(\Pi_{\textbf{k}}^S)_{ij} = \text{diag}(\frac{1}{3}, \frac{1}{3}, -\frac{2}{3})\Pi_{\textbf{k}}$. From Eq.~(\ref{Tparam}), the scalar perturbation to the spatial part of the stress tensor in a spatially flat slicing will be $(\delta T_{\textbf{k}}^S)_{ij} = -\text{diag}( c_s^2 - 2c_v^2, c_s^2 - 2c_v^2, c_s^2)\delta\rho_{\textbf{k}}$, which is inherently anisotropic, having a different pressure in directions parallel and perpendicular to the direction of propagation.

Instead of considering perturbations about a FRW spacetime, we now examine the behaviour of an unperturbed Bianchi spacetime, which has the defining properties of being homogeneous and in general anisotropic. We concentrate on the Bianchi type I spacetime that has the metric
\begin{equation}
ds^2 = dt^2 - a_x(t)^2 dx^2 - a_y(t)^2 dy^2 - a_z(t)^2 dz^2
\end{equation}
where $a_x$, $a_y$, and $a_z$ are directional scale factors. The properties of nearly isotropic Bianchi spacetimes were detailed in Ref.~\cite{bianchi}, which treated their departure from isotropy as a linear perturbation. After smoothing on superhorizon scales, the metric perturbation $E_{ij}$ in Eq.~(\ref{metric}) from perturbing about a flat FRW spacetime is simply the symmetric trace-free tensor characterizing the anisotropy of a nearly isotropic Bianchi~I spacetime to linear order.\footnote{See Eq.~(59) of Ref.~\cite{bianchi}. Also note that on superhorizon scales, by making the substitution $\sigma\rightarrow\dot{E}$, Eq.~(\ref{shearevolution}) is approximately Eq.~(39) in Ref.~\cite{bianchi}.} For example, consider the mode $\textbf{k}=(0,0,k)$. After smoothing, $E_{ij}$ will be approximately uniform in a local patch of the Universe. In a spatially flat gauge where $h=2\nabla^2E$, if we denote the average value of $h$ in this patch as $\bar{h}$, then a scalar mode with this wavevector will evolve approximately as a Bianchi~I spacetime with $a_x=a_y=a$ and $a_z=a+\bar{h}$. A tensor mode with `plus' polarization, with an average value of $\bar{h}_+$ in this patch, would evolve with directional scale factors $a_x=a+\bar{h}_+$, $a_y=a-\bar{h}_+$, and $a_z=a$.

For a single mode, we can absorb the shear on superhorizon scales into the background spacetime by perturbing about a Bianchi I spacetime instead of a flat FRW (which is the isotropic special case of Bianchi I). In this case, after smoothing on superhorizon scales, perturbations would again evolve according to an unperturbed metric, but in general would be of type Bianchi I, not FRW. In the standard inflationary scenario, the shear is negligible on superhorizon scales so $E_{ij}$ is approximately constant on these scales and can be removed from the metric by a simple coordinate redefinition, leaving the isotropic special case of our spacetime.

Although a single mode is formally anisotropic, in the case of inflation, modes are excited on a wide range of scales in all directions. We assume that initial perturbations are drawn from an isotropic Gaussian distribution and that expectation values will be isotropic and therefore continue to examine the perturbations in the metric in Eq.~(\ref{basicmetric}) in which perturbations are taken about an isotropic FRW spacetime.\footnote{Although we take the expectation values over a single realization to be isotropic, in principle, a residual net anisotropy might persist. The persistence of anisotropic geometries in this context was recently studied in Ref.~\cite{anisotropicSolid}. We set aside here questions pertaining to the precise size and impact of sustained anisotropies and only assume that corrections to the evolution of linear perturbations in an isotropic background appear at higher order.}

\section{Inflation}
\label{inflationSection}

We now apply the results of the previous sections to the case where inflation is driven by an elastic solid. We divide the analysis into two parts: the simple case with constant sound speeds and equation of state and the case where they are varying in time. We then consider the more specialized case where the sound speeds and equation of state slowly vary with time.

\subsection{Inflation with Constant Sound Speeds and Equation of State}
\label{constSpeedSec}

With constant equation of state, $\mathcal{H}$ can easily be solved from Eq.~(\ref{HDE}) with an appropriate integration constant as
\begin{equation}
\mathcal{H} = \frac{2}{(1+3w)\eta}
\label{Hconst}
\end{equation}
From Eq.~(\ref{meff}), we see that in this case the effective mass for the scalar modes becomes
\begin{equation}
 m^2_{\text{eff},S}(\eta) = - \frac{2-6w-24c_v^2(1+w)}{(1+3w)^2\eta^2}
\label{meffconst}
\end{equation}
The general solution for the mode function $\chi_{\textbf{k}}(\eta)$ can now easily be found from Eq.~(\ref{chieom}) as
\begin{equation}
	\chi_k = \sqrt{\frac{\pi|\eta|}{2}} \left[ C_1 H^{(1)}_{\nu}(c_s k|\eta|)+C_2 H^{(2)}_{\nu}(c_s k|\eta|) \right]
\label{chisol}
\end{equation}
where $H^{(1)}_{\nu}$ and $H^{(2)}_{\nu}$ are the Hankel functions of the first and second kind, $C_1$ and $C_2$ are integration constants and the index $\nu$ is
\begin{equation}
	\nu = \frac{1}{2}\sqrt{1+4\left(\frac{2-6w-24c_v^2(1+w)}{(1+3w)^2}\right)}
\end{equation}
When a mode is well within the horizon with $c_s k|\eta|\gg 1$, the mode function can be approximated by
\begin{equation}
	\chi_k \approx \frac{1}{\sqrt{c_s k}}\left(C_1e^{ic_s k\eta}+C_2e^{-ic_s k\eta}\right)
\label{chipast}
\end{equation}
which is the solution for Minkowski space. We initialize the mode by assuming that it is in its lowest energy state when it is well within the horizon, with mode function
\begin{equation}
	\chi_k \approx \frac{1}{\sqrt{c_s k}}e^{ic_s k\eta}
\label{chiinit}
\end{equation}
With these constants of integration, the mode function becomes
\begin{equation}
	\chi_k = \sqrt{\frac{\pi|\eta|}{2}}H^{(2)}_{\nu}(c_s k|\eta|)
\label{chieom3}
\end{equation}
When the mode is far outside the horizon with $c_s k|\eta|\ll 1$, the mode function can be approximated as
\begin{equation}
	\chi_k \approx \sqrt{\frac{\pi|\eta|}{2}} \frac{i\Gamma(\nu)}{\pi}\left(\frac{c_sk|\eta|}{2}\right)^{-\nu}
\label{chilimit}
\end{equation}
where $\Gamma(\nu)$ is the gamma function. With the evolution of the mode function $\chi_{\textbf{k}}$ for modes well outside the horizon, we find the power spectrum for $R$ to be
\begin{equation}
	\mathcal{P}_R(k) \approx \frac{c_s^{2(1-\nu)} \Gamma^2(\nu)4^{\nu}}{8\pi^3(1+w)}\frac{l^2 k^{3-2\nu}|\eta|^{1-2\nu}}{a^2}
\label{scalarpower}
\end{equation}
Since for constant equation of state the scale factor evolves as $a\propto |\eta|^{\frac{2}{1+3w}}$, we do indeed see that $R_k$ evolves with time when the mode is on superhorizon scales if $c_v$ is nonzero. Using Eq.~(\ref{Hconst}), we can write the power spectrum in terms of the Hubble parameter as
\begin{equation}
	\mathcal{P}_R(k) \approx \frac{c_s^{2(1-\nu)} \Gamma^2(\nu)}{4\pi^3(1+w)|1+3w|^{1-2\nu}} l^2 (k/a)^{3-2\nu} H^{-1+2\nu}
\end{equation}

Although modes evolve on superhorizon scales, all modes well outside the horizon share the same time evolution. In other words, the presence of a superhorizon evolution will not affect the relative scale dependence of modes on superhorizon scales. Thus, we can calculate quantities like the scalar spectral index $n_s = 1 + d\text{ln}\mathcal{P}_{R}/d\text{ln}k$ and its running using the same methods that are used in the case where the superhorizon evolution is small.

For constant sound speeds and equation of state, the scalar spectral index is $n_s = 4-2\nu$. The necessary restrictions of $-1 < w <-1/3$, $0\leq c_v^2\leq1$, and $0<c_s^2\leq1$ imply that $n_s$ is bound from below by one. Thus, the scalar power spectrum for this case can only have a blue tilt, which has been ruled out to a high degree of likelihood \cite{wmap9}. If $w$ is near $-1$, we can see from Eq.~(\ref{meffconst}) that the effect of the shear stress on $m^2_{\text{eff},S}$ will be small and a nearly scale-invariant spectrum will be produced, as is the case in many models of inflation.

It is interesting to note that $c_s$ near zero ($w$ near $-\frac{4}{3}c_v^2$) also produces a nearly scale-invariant two-point spectrum. In this case, the $w$ dependence of the $-z''/z$ and $4c_v^2\beta$ terms in $m^2_{\text{eff},S}$ cancels with one another so that a nearly scale-invariant spectrum can be produced for values of $w$ far from $-1$ (but still bounded by $-1<w<-1/3$). This result is not possible in standard inflationary models since the $4c_v^2\beta$ term is absent in these cases, so the $w$ dependence of $m^2_{\text{eff},S}$ remains important.

As discussed in Ref.~\cite{baumann}, inflationary scenarios that produce a nearly scale-invariant two-point function with a background in the far from de Sitter regime generically do not have nearly scale-invariant higher-point correlations provided the perturbations are adiabatic in the sense of Ref.~\cite{weinberg} --- which requires the anisotropic stress to be negligible on large scales. However, the elastic solid model we describe here requires non-negligible anisotropic stress on large scales for a consistent description of linear perturbations.  We leave the interesting question of higher-point correlation functions in far from de Sitter accelerating elastic solid models for future work.

Additionally, when $w$ is far from $-1$, one must take care that an adequate number of $e$-folds of inflation can occur as the energy density and horizon size may evolve significantly during inflation. This can alter the minimum number of $e$-folds required to solve the `horizon problem'. An upper bound on the number of $e$-folds of inflation will be set by putting bounds on the energy density, set on the lower end by the reheat temperature and on the higher end by a high-energy limit (see Appendix~\ref{efolds} for further details). Requiring inflation to start below the Planck scale and end with temperatures above $\sim$10's of MeV puts an upper bound on the equation of state of $w\lesssim-2/5$ when a nearly scale-invariant spectrum is achieved. However, as will be discussed in Section~\ref{reheating},  a power spectrum amplitude compatible with current observations requires either very small values of $c_s$ or supra-Planckian densities for values of $w$ extremely far from $-1$.

\begin{figure}[H]
  \centering
    \includegraphics[width=\linewidth]{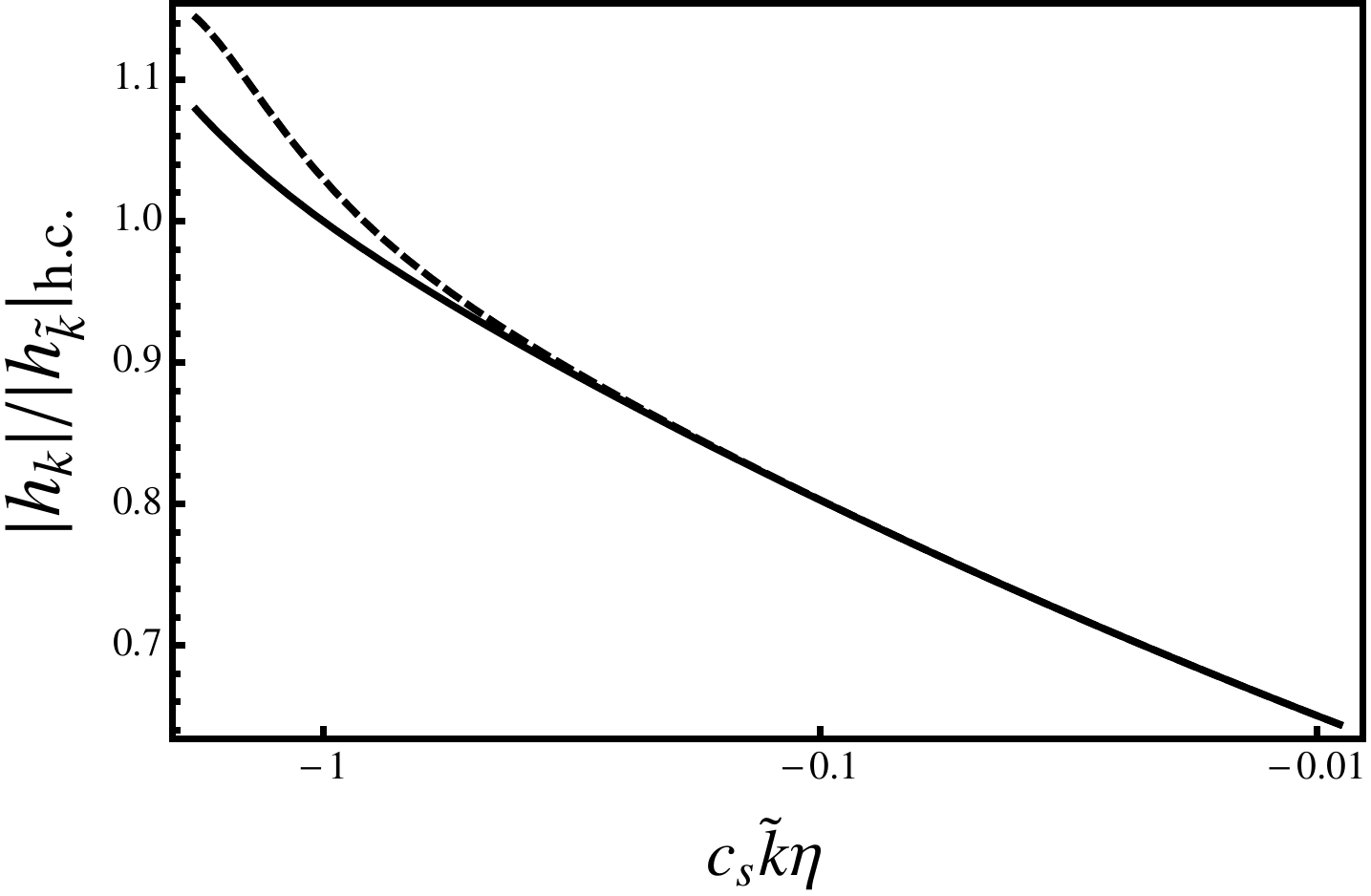}
	\caption{Evolution of $h$ modes in the $\psi=B=0$ gauge for $w=-0.9$ and $c_v^2=0.8$ (note that a plot of $E_k$ would look identical as $h_{\textbf{k}}=-2 E_{\textbf{k}}$ in this gauge). The solid and dashed lines show the evolution for modes with $|\textbf{k}|=\tilde{k}$ and $|\textbf{k}|=2\tilde{k}$, respectively, and the subscript ${\rm{h.c.}}$ denotes horizon crossing. Initial amplitudes of the perturbations are chosen so that the modes coincide when both are on superhorizon scales.}
\label{E_plt}
\end{figure}

Returning to the discussion of Section~\ref{super}, for constant sound speeds and equation of state with $c_v\neq0$, the superhorizon modes of $h$ in the $\psi=B=0$ gauge (as well as $E$ since $h=2\nabla^2E$ in this gauge) evolve as
\begin{equation}
	h_{\textbf{k}}' \propto A_{\textbf{k}} k^{-\nu} |\eta|^{-\frac{5+3w}{2+6w} - \nu }
\end{equation}
where the factor $A_{\textbf{k}}$ determines the initial amplitude of $h_{\textbf{k}}' $ for a particular mode. If we scale the wavevector of the mode as $\textbf{k} \rightarrow \alpha \textbf{k}$ for some constant $\alpha$, the same late-time evolution of the superhorizon mode can remain unchanged by simultaneously scaling $A_{\textbf{k}} \rightarrow \alpha^{\nu} A_{\textbf{k}}$, an example of which can be seen in Fig.~\ref{E_plt}. As such, we cannot determine which particular mode a superhorizon sized anisotropy originated from.

\subsection{Non-Constant Sound Speeds and Equation of State}
\label{nonconstspeeds}

Since having the sound speeds and equation of state perfectly constant cannot result in a red-tilted scalar spectrum, we would like to examine if adding a time dependence can result in a red-tilted scalar spectrum. It will prove useful to introduce an alternative time variable $q$, defined by
\begin{equation}
	q(\eta) \equiv -\int^{\eta} c_s(\tilde{\eta}) d\tilde{\eta}
\label{sigmadef}
\end{equation}
and a new field $y_{\textbf{k}}\equiv \sqrt{c_s}\chi_{\textbf{k}}$, so that the equation of motion for the mode function in Eq.~(\ref{chieom}) becomes
\begin{equation}
	y_{\textbf{k},qq} + \left[k^2+\tilde{m}^2_{\text{eff},S}\right] y_{\textbf{k}} = 0
\label{yeom}
\end{equation}
where $,_q=d/dq$ and
\begin{equation}
	\tilde{m}^2_{\text{eff},S} \equiv \frac{m^2_{\text{eff},S}}{c_s^2}-\frac{(\sqrt{c_s}),_{qq}}{\sqrt{c_s}}
\end{equation}
The advantage of this change of variables is that the squared sound speed $c_s^2$ does not appear in front of the $k^2$ term in Eq.~(\ref{yeom}), as it does in Eq.~(\ref{chieom}), so that the same methods used for solving for the equation of motion of $\chi_{\textbf{k}}$ in the case where $c_s$ is constant can be used to solve for the new mode function $y_{\textbf{k}}$ as a function of the new time variable $q$.

In the previous section, the mode function was easily solved because $m^2_{\text{eff},S}$ was inversely proportional to $\eta^2$. Accordingly, we reparametrize $\tilde{m}^2_{\text{eff},S}$ by $\mathcal{B}_S\equiv -q^2\tilde{m}^2_{\text{eff},S}$, so that the equation of motion for $y_{\textbf{k}}$ becomes
\begin{equation}
	y_{\textbf{k},qq} + \left[k^2 -\frac{\mathcal{B}_S}{q^2} \right] y_{\textbf{k}} = 0
\label{yeom2}
\end{equation}
To obtain a solution where the running of $n_s$ is small, we consider solutions where $\mathcal{B}_S$ is nearly constant, in which case Eq.~(\ref{yeom2}) would have solution
\begin{equation}
	y_k(q) \approx \sqrt{\frac{\pi q}{2}} \left[ C_1 H^{(1)}_{\gamma_S}(kq)+C_2 H^{(2)}_{\gamma_S}(kq) \right]
\label{ysoln}
\end{equation}
where the index $\gamma_S$ is
\begin{equation}
	\gamma_S \equiv \frac{1}{2}\sqrt{1+4\mathcal{B}_S}
\label{gammadef}
\end{equation}
As with the previous case, we choose the integration constants $C_1$ and $C_2$ to select the lowest energy state when $kq\gg1$. This coincides with the asymptotic solution in Eq.~(\ref{chiinit}) if the change of $c_s$ with time is small at some early time when all modes of interest are well within the horizon. Writing the normalization condition in Eq.~(\ref{norm}) in terms of the mode function $y_{\textbf{k}}$ and the time variable $q$ gives
\begin{equation}
	(y_{\textbf{k}}),_{q}y_{\textbf{k}}^* - y_{\textbf{k}} (y_{\textbf{k}})^*,_{q} = -2i
\label{ynorm}
\end{equation}
With these choices, the mode function $y_{\textbf{k}}$ evolves as
\begin{equation}
		y_k(q) = \sqrt{\frac{\pi q}{2}}H_{\gamma_S}^{(1)}(kq)
\label{yHankel}
\end{equation}
When dealing with the time variable $q$, $k|q|\sim 1$ does not necessarily imply $c_s k|\eta|\sim 1$. However, in the case when $c_s$ varies slowly in time, as is considered in Section~\ref{slowroll}, then $k|q|\sim 1$ when $c_s k|\eta|\sim 1$, in which case there will be no confusion about what is meant by a mode crossing the horizon.

Analogously with the previous section, on superhorizon scales, the mode function $y_k$ is approximately equal to 
\begin{equation}
	y_k(q) \approx -\sqrt{\frac{\pi q}{2}} \frac{i\Gamma(\gamma_S)}{\pi} \left(\frac{kq}{2}\right)^{-\gamma_S}
\label{lateapprox}
\end{equation}
at which time the power spectrum for $R$ becomes
\begin{equation}
	\mathcal{P}_R(k) \approx \frac{c_s \Gamma^2(\gamma_S) 4^{\gamma_S}}{8\pi^3(1+w)}\frac{l^2 k^{3-2\gamma_S}q^{1-2\gamma_S}}{a^2}
\label{Py}
\end{equation}
which implies that the scalar spectral index $n_s$ is now given by
\begin{equation}
	n_s = 4-2\gamma_S
\label{nsgamma}
\end{equation}
It is trivial to check that in the case that the sound speeds and equation of state are constant, the above equations simplify to those given in the previous section. However, if we can find time-varying sound speeds and/or equation of state such that $\mathcal{B}_S$ is approximately constant, the bounds on the index $\gamma_S$ may be extended to include red-tilted scalar spectra.

\subsection{Slowly Varying Sound Speeds and Equation of State}
\label{slowroll}

From the above considerations, we wish to find a parametrization of the sound speeds and equation of state that allows for a small variation in time in such a way that results in $\mathcal{B}_S$ being approximately constant and allows the scalar spectral index to be less than one.

We will use the variable $\epsilon$, which coincides with the slow-roll variable from standard inflationary scenarios, defined by
\begin{equation}
\epsilon \equiv -\frac{\dot{H}}{H^2} = 1-\frac{\mathcal{H}'}{\mathcal{H}^2}
\end{equation}
so that $w=-1+\frac{2}{3}\epsilon$. In this context, $\epsilon$ simply parametrizes the departure of $w$ from -1. We parametrize $\epsilon$ as $\epsilon(\eta) = \epsilon_0 + f_{\epsilon}(\eta)$ for some slowly varying function $f_{\epsilon}(\eta)$. We write the time dependence of $f_{\epsilon}$ as
\begin{equation}
\frac{d\text{ln}f_{\epsilon}}{d\text{ln}\eta} = -\tau_{\epsilon}
\end{equation}
If we assume that $|\tau_{\epsilon}|\ll1$, then $\epsilon(\eta)$ will be given by
\begin{equation}
\epsilon(\eta) \approx \epsilon_0 + \epsilon_1(\eta/\eta_*)^{-\tau_{\epsilon}}
\label{epsilonparam}
\end{equation}
for some reference time $\eta_*$. With this time dependence, $w$ slowly varies near $-1+\frac{2}{3}\epsilon_0$ for some time, but at some later time, it will evolve at a more rapid pace. We will choose the reference time $\eta_*$ to be the end of inflation to ensure that the time dependence of $w$ is small during inflation. We also allow $c_s$ to vary in time and use a parametrization analogous to the one used for $\epsilon$, so that
\begin{equation}
c_s(\eta) \approx c_{s0} + c_{s1}(\eta/\eta_*)^{-\tau_s}
\label{csparam}
\end{equation}
where $|\tau_s|\ll1$. With this parameterization, our new time variable $q$ is
\begin{equation}
q(\eta) = -\eta\left[  c_{s0} + \frac{c_{s1}(\eta/\eta_*)^{-\tau_s}}{1-\tau_s} \right]
\end{equation}

The requirement that $\mathcal{B}_S$ be approximately constant for $\eta\leq\eta_*$ is met so long as both $\tau_{\epsilon}$ and $\tau_s$ are sufficiently small. Note that solutions where $w$ departs significantly from $-1$ are valid since $\epsilon_0$ is not required to be small. If we desire $w$ to stay close to $-1$, we would add the restrictions $|\epsilon_0|\ll1$ and $|\epsilon_1|\ll1$. To obtain a slightly red-tilted scalar spectrum, we will want the sound speeds and equation of state to evolve near constant values that result in a nearly scale-invariant (blue-tilted) scalar spectrum. Therefore, from Section~\ref{constSpeedSec}, we will want to consider solutions where at least one of $\epsilon_0$ and $c_{s0}$ are small.

For the rest of this paper, we restrict ourselves to cases where $|\epsilon_1|\ll1$, in which case Eq.~(\ref{HDE}) can be used to solve for $\mathcal{H}$ and subsequently the scale factor $a$, which to linear order in $\tau_{\epsilon}$ and $\epsilon_1$ is
\begin{equation}
\mathcal{H} \approx -\frac{1-\epsilon_0+\epsilon_1}{\eta(1-\epsilon_0)^2} \quad\quad a \approx a_*(\eta/\eta_*)^{-\frac{1-\epsilon_0+\epsilon_1}{(1-\epsilon_0)^2} }
\end{equation}
where $a_*$ is the value of the scale factor at $\eta=\eta_*$. In the case where $\epsilon_0=c_{s0}=0$, $\mathcal{B}_S$ to first order in our small parameters is found to be
\begin{equation}
\mathcal{B}_S \approx 2 +\frac{15}{2}\tau_s + \frac{3}{2}\tau_{\epsilon} - 3c_{s1}^2\epsilon_1
\end{equation}
and the scalar spectral index $n_s$ becomes
\begin{equation}
n_s \approx 1 - 5\tau_s - \tau_{\epsilon} + 2c_{s1}^2\epsilon_1
\label{nsapprox}
\end{equation}
from which we see can yield a red-tilted scalar spectrum (see the first three rows of Table~\ref{exampleVals} for examples). Note that although $c_{s}\rightarrow 0$ if $c_{s0}=0$ in the limit where $\eta\rightarrow -\infty$, $c_s$ at the beginning of inflation will not be significantly different from its value at the end of inflation ($c_s(\eta_*)=c_{s1}$) as long as $|\tau_s|\ll1$ and the number of $e$-folds of inflation is modest (i.e. 100's of $e$-folds).

As previously stated, we can still estimate the running of $n_s$ by conventional means where relevant quantities are evaluated at horizon crossing. As horizon crossing occurs when $c_s k|\eta| \sim 1$, at horizon crossing $d\text{ln}k \sim -(\eta^{-1} + c_s'/c_s)^{-1} d\eta$, so that 
\begin{equation}
\frac{d n_s}{d\text{ln}k} \sim \left( \frac{1}{\eta} + \frac{c_s'}{c_s} \right)^{-1} \frac{d}{d\eta} \sqrt{1+4\mathcal{B}_S} \bigg|_{c_s k|\eta| \sim 1}
\end{equation}
For $\epsilon_0=c_{s0}=0$, the running to second order in our small parameters is
\begin{equation}
\frac{d n_s}{d\text{ln}k} \sim 2 c_{s1}^2 \epsilon_1  (\tau_{\epsilon} + 2\tau_s)
\end{equation} 
from which we see that the running of $n_s$ vanishes at first order.

As in the constant sound speeds and equation of state case, we find that we are not restricted to very small values of $\epsilon_0$ and $c_{s0}$, although for brevity we will not explicitly write out $n_s$ and its running and instead illustrate through numerical examples. In fact, we can formally find solutions with $w$ varying slowly near values up to $-1/3$, corresponding to values of $\epsilon_0$ just below 1, that result in a slightly red-tilted scalar spectrum with a small running of its spectral index, although, as previously mentioned, achieving the necessary number of $e$-folds of inflation becomes increasingly challenging for values of $w$ very far from $-1$.

As an example, choosing $\epsilon_0=1/4$ so that $w$ varies close to $-5/6$ and taking the other parameters to have the values listed in the fourth row of Table \ref{exampleVals}, we obtain a scalar spectral index $n_s \approx 0.96$ and running $dn_s/d\text{ln}k \sim -10^{-5}$. Another example where $w$ varies near $-2/3$ is given in the fifth row of Table~\ref{exampleVals}. Note that the reheat temperature used in this example is significantly lower compared to the other examples listed in the table to allow for the required number of $e$-folds of inflation (see Appendix~\ref{efolds} for details). Choosing $c_{s0}$ to be nonzero instead of $\epsilon_0$, with $c_{s0}=0.15$ and the values in the last row of Table \ref{exampleVals} yields a scalar spectral index of $n_s\approx 0.96$ and running $dn_s/d\text{ln}k \sim -10^{-3}$.

We conclude this section by writing the power spectrum for $R$ at the end of inflation for superhorizon modes, which from Eq.~(\ref{Py}), is
\begin{multline}
\mathcal{P}_R(k) \approx \frac{3 \Gamma^2(\gamma_S)(c_{s0}+c_{s1})}{8\pi^3(\epsilon_0+\epsilon_1)} \\ \times \left[  \frac{c_{s0}(1-\epsilon_0+\epsilon_1)+c_{s1}(1+\epsilon_1+\tau_s-\epsilon_0(1+\tau_s))}{2(1-\epsilon_0)^2} \right]^{1-2\gamma_S} \\ \times l^2 (k/a_*)^{3-2\gamma_S} H_*^{-1+2\gamma_S}
\label{PRinf}
\end{multline}
where the subscript $*$ denotes evaluation at $\eta=\eta_*$. Since modes can evolve on superhorizon scales, $\mathcal{P}_R(k)$ at the end of reheating may be different from the expression given above. This concern will be addressed in Section~\ref{reheating}.

\begin{table*}[!ht]
\begin{tabular}{| c | c | c | c | c | c | c | c | c | c | c | c |}
\hline
$c_{s0}$ & $c_{s1}$ & $\epsilon_0$ & $\epsilon_1$ & $\tau_s$ & $\tau_{\epsilon}$ & $T_{RH} (\text{GeV})$ & $n_s$ & $10^9 A_{\zeta}$ & $r(k_p)$ & $n_T$ & $dn_s/d\text{ln}k$ \\ \hline
0 & 0.01 & 0 & 0.01 & 0.005 & 0.015 & $1.55\times10^{13}$ & 0.96 & 2.43 & $\sim 10^{-12}$ & $-3.1\times 10^{-4}$ & $\sim -10^{-5}$  \\ \hline
0 & 0.8 & 0 & 0.01 & 0.0086 & 0.01 & $3.69\times 10^{15}$ & 0.96 & 2.43 & 0.002 & 0.013 & $\sim10^{-4}$  \\ \hline
0 & 0.3 & 0 & 0.001 & 0.005 & 0.015 & $6.06\times10^{14}$ & 0.96 & 2.43 & $\sim 10^{-6}$ & $1.5\times 10^{-4}$ & $\sim 10^{-6}$  \\ \hline
0 & 0.001 & 0.25 & 0.01 & 0.0078 & 0.0078 & $3.05\times 10^{12}$ & 0.96 & 2.43 & $\sim 10^{-15}$ & $-0.0002$ & $\sim -10^{-5}$   \\ \hline
0 & $10^{-8}$ & 0.5 & 0.01 & 0.008 & 0.001 & $3.84 \times 10^6$ & 0.96 & 2.43 & $\sim 10^{-40}$ & $-4.1\times 10^{-5}$ & $\sim -10^{-7}$ \\ \hline
0.15 & 0.05 & 0 & 0.01 & 0.024 & 0.01 & $1.79\times 10^{14}$ & 0.96 & 2.43 & $\sim 10^{-6}$ & $6.1\times 10^{-4}$ & $\sim -10^{-3}$   \\ \hline
\end{tabular}
\caption{Examples of choice of parameters for slowly varying sound speeds and equation of state. For cases where $c_{s0}=\epsilon_0=0$, $n_s$, $n_T$, and $r$ are calculated directly from Eqs.~(\ref{nsapprox}), (\ref{nT}), and (\ref{rapprox}), respectively. In examples where either $\epsilon_0\neq 0$ or $c_{s0}\neq 0$, $n_s$ and $n_T$ are calculated by first computing $\mathcal{B}_S$ or $\mathcal{B}_T$, as defined in Sections~\ref{nonconstspeeds} and \ref{gravwaves}, respectively. The power spectrum for $\zeta$ has been parametrized as $P_{\zeta}(k)=A_{\zeta}(k/k_p)^{n_s-1}$ where $k_p=0.002 \, \text{Mpc}^{-1}$. In calculating $A_{\zeta}$ and $r$, the model of a rapid decay into radiation was assumed to end inflation and reheat the Universe. For cases with $c_{s0}=\epsilon_0=0$, $A_{\zeta}$ and $r$ are calculated using Eqs.~(\ref{Azeta}) and (\ref{rapprox}), respectively. For examples with either $\epsilon_0\neq 0$ or $c_{s0}\neq 0$,  $A_{\zeta}$ and $r$ are calculated using the power spectra given in Eqs.~(\ref{PRinf}) and (\ref{PhH}), respectively, along with the relevant equations found in Section~\ref{reheating} to relate $R$, $h^T$ just before the end of inflation to $\zeta$, $h^T$ following the end of inflation. All examples use $g_{RH}=106.75$.}
\label{exampleVals}
\end{table*}

\section{Gravitational Waves}
\label{gravwaves}

To find the amplitude of gravitational waves produced during inflation, we need solve for the equation of motion of the mode function $X_{\textbf{k}}$ in Eq.~(\ref{tensoreom}), which can be solved in an analogous manner to $\chi_{\textbf{k}}$ in the scalar case. This task will be relatively easy compared to the scalar perturbations, since the tensor perturbations travel with a sound speed equal to unity so there is no need to switch time variables in order to solve the differential equation for $X_{\textbf{k}}$. Analogous to the scalar case, for the tensor modes we can define
\begin{equation}
	\mathcal{B}_T \equiv -\eta^2 m^{2}_{\text{eff},T}
\end{equation}
Using the parametrizations of the sound speeds and equation of state in Section~\ref{slowroll}, $\mathcal{B}_T$ to first order in $\tau_s$, $\tau_{\epsilon}$, and $\epsilon_1$ with $\epsilon_0=c_{s0}=0$ is
\begin{equation}
	\mathcal{B}_T \approx 2 - 3c_{s1}^2\epsilon_1
\end{equation}
Since $\mathcal{B}_T$ is constant to linear order in our small parameters, the tensor mode function in Eq.~(\ref{tensoreom}) has solution
\begin{equation}
	X_k \approx \sqrt{\frac{\pi|\eta|}{2}}\left[ C_1 H^{(1)}_{\gamma_T}(k|\eta|) + C_2 H^{(2)}_{\gamma_T}(k|\eta|) \right]
\end{equation}
where $C_1$ and $C_2$ are new constants of integration and the index $\gamma_T$ is
\begin{equation}
	\gamma_T = \frac{1}{2}\sqrt{1+4\mathcal{B}_T}
\end{equation}
The choice of $C_1=0$ and $C_2=1$ satisfies the normalization condition of Eq.~(\ref{normtensor}) and selects the lowest energy state when the mode is well within the horizon. When modes are well outside the horizon, the mode function can be approximated as
\begin{equation}
	X_k \approx \sqrt{\frac{\pi|\eta|}{2}} \frac{i\Gamma(\gamma_T)}{\pi}\left(\frac{k|\eta|}{2}\right)^{-\gamma_T}
\label{Xlimit}
\end{equation}
Using this approximation for the mode function with Eq.~(\ref{Ph}) yields the power spectrum for gravitational waves on superhorizon scales
\begin{equation}
	\mathcal{P}_T (k) \approx \frac{6 \times 2^{2\gamma_T-1}l^2 \Gamma^2(\gamma_T)}{\pi^3} \frac{k^{3-2\gamma_T}|\eta|^{1-2\gamma_T}}{a^2}
\label{Phn}
\end{equation}
When $\epsilon_0=c_{s0}=0$, the tensor spectral index $n_T=d\text{ln}\mathcal{P}_T/d\text{ln}k$ is 
\begin{equation}
	n_T \approx 2 c_{s1}^2\epsilon_1
\label{nT}
\end{equation}
At the end of inflation, the tensor power spectrum can be written as
\begin{multline}
	\mathcal{P}_T(k) \approx \frac{6 \Gamma^2(\gamma_T)}{\pi^3} \left( \frac{1-\epsilon_0+\epsilon_1}{2(1-\epsilon_0)^2} \right)^{1-2\gamma_T} \\ \times l^2 (k/a_*)^{3-2\gamma_T} H_*^{-1+2\gamma_T}
\label{PhH}
\end{multline}

\section{End of Inflation and Reheating}
\label{reheating}

Unlike in cases where the superhorizon evolution of modes is small, the details of the end of inflation will affect the amplitude of modes on superhorizon scales, although the time evolution of all superhorizon modes will be the same. There are many possibilities for ending inflation and reheating within this model. One possibility is once all modes of interest are on superhorizon scales having $w$ increase rapidly so that it surpasses $-1/3$ and the Universe stops inflating. At some later point, the solid can lose its rigidity, at which point the superhorizon evolution will be small and the details of reheating will not affect modes on superhorizon scales. Another possibility, which we will examine in more detail, is that inflation ends with the decay of the elastic solid. 

In general, $\zeta_k$ and $R_k$ will not be equal during inflation on superhorizon scales. We can easily find the relationship between $\zeta$ and $R$ by using Eqs.~(\ref{pieqn2}) and (\ref{Piscalar}), which is
\begin{equation}
\zeta = \frac{d P/d\rho}{c_s^2}R - \frac{1}{3c_s^2\mathcal{H}}R'
\label{zetaR}
\end{equation}
from which we see that in general, even on superhorizon scales,
 $\zeta_k \neq R_k$. After inflation ends and the rigidity vanishes, the superhorizon evolution will be small and $\zeta_k \approx R_k$, but the change in $\zeta_k$ and  $R_k$ must be tracked through the transition that ends inflation.
 
We now consider the case where the elastic solid rapidly decays into a perfect fluid to end inflation. Following Ref.~\cite{interactingFluids}, we can write the total stress-energy tensor as
\begin{equation}
T^{\mu\nu} = \sum_{\alpha} T^{\mu\nu}_{(\alpha)}
\end{equation}
where $T^{\mu\nu}_{(\alpha)}$ is the stress-energy tensor of component $\alpha=\{e,f\}$, with $e$ and $f$ denoting the elastic solid and perfect fluid decay product, respectively. For this analysis, it will be more convenient to use the coordinate time $t$ instead of the conformal time. While the local energy-momentum transfer 4-vector $Q^{\nu}_{(\alpha)}$ for each species can be nonzero, so that
\begin{equation}
\nabla_{\mu} T^{\mu\nu}_{(\alpha)} = Q^{\nu}_{(\alpha)}
\label{Talpha}
\end{equation}
we must have $\sum_{\alpha} Q^{\nu}_{(\alpha)} = 0$ so that the total stress-energy tensor is covariantly conserved.

For the scalar perturbations, we can define a $\zeta_{\alpha}$ variable for each substance $\alpha$, defined by
\begin{equation}
\zeta_{\alpha} \equiv \psi + H \frac{\delta\rho_{\alpha}}{\dot{\rho}_{\alpha}}
\label{zetaalpha}
\end{equation}
that is related to the total $\zeta$ by
\begin{equation}
\zeta = \sum_{\alpha} \frac{\dot{\rho}_{\alpha}}{\dot{\rho}} \zeta_{\alpha}
\end{equation}
Similarly, the variable $R_{\alpha}$ for each substance, defined by
\begin{equation}
R_{\alpha} = \mathcal{H} v_{\alpha} + \psi
\label{Ralpha}
\end{equation}
is related to the total $R$ by
\begin{equation}
R = \sum_{\alpha} \frac{\rho_{\alpha} + P_{\alpha}}{\rho + P} R_{\alpha}
\end{equation}
The total energy density $\rho$ and pressure $P$ are given by $\rho = \sum_{\alpha} \rho_{\alpha}$ and $P = \sum_{\alpha} P_{\alpha}$, while the entropy perturbation between substances $\alpha$ and $\beta$ is
\begin{equation}
\mathcal{S}_{\alpha\beta} \equiv 3(\zeta_{\alpha} - \zeta_{\beta})
\end{equation}
and its relative velocity perturbation is given by
\begin{equation}
v_{\alpha\beta} \equiv v_{\alpha} - v_{\beta} = \frac{R_{\alpha}-R_{\beta}}{\mathcal{H}}
\end{equation}

For the decay of the elastic solid to a perfect fluid, the energy-momentum transfer is given by
\begin{equation}
Q^{\nu}_e = -Q^{\nu}_f = - \Gamma g^{\nu\beta}u_{\beta}\rho_e(1+w_e)
\label{Qdecay}
\end{equation}
where $u_{\beta}$ is the total velocity 4-vector of the elastic solid and the perfect fluid and $\Gamma$ is the decay rate of the elastic solid into the fluid (not to be confused with the gamma function used previously).

In the current case, where a single substance is decaying into another single substance, we expect that entropy perturbations will not be generated in the decay. In general, the evolution of the entropy perturbation between the elastic solid and the fluid is given by
\begin{multline}
(\dot{\mathcal{S}}_{ef})_k = \left[ \frac{\dot{Q}_e}{\dot{\rho}_f} + \frac{Q_e}{2\rho}\left( \frac{\dot{\rho}_f}{\dot{\rho}_e} - \frac{\dot{\rho}_e}{\dot{\rho}_f}     \right) \right] (\mathcal{S}_{ef})_k \\ + \frac{k^2}{a^2H} \left[ \left(1-\frac{Q_f}{\dot{\rho}_f}\right)(R_f)_k  - \left(1-\frac{Q_e}{\dot{\rho}_e}\right)(R_e)_k  \right]
\label{SDE}
\end{multline}
where $Q_{\alpha}=Q_{(\alpha)}^0$ is the background value of the time component of the energy-momentum-transfer 4-vector. We refer to Appendix~\ref{decay} for the explicit form of the background and perturbation equations used to derive this relation. We can see that if on superhorizon scales the entropy perturbation vanishes at some time, the entropy perturbation will stay approximately constant past this time. If the decay is rapid $(\Gamma\gg H)$, then any preexisting entropy perturbations will quickly be driven to zero at the very beginning of the decay. Therefore, for times of interest, we set the entropy perturbation to zero. With no entropy perturbation, $\zeta_e = \zeta_f$ and $\zeta$ will be continuous across the decay, when it changes from $\zeta = \zeta_e$ before the decay to $\zeta= \zeta_f$ after the decay. On the other hand, we do not expect the relative velocity perturbation to be zero during the decay, so in general $R$ will change rapidly during the decay as it goes from $R=R_e$ before the decay to $R=R_f\approx \zeta_f$ on superhorizon scales after the decay. In this light, we will follow $\zeta$ instead of $R$ from the end of inflation into radiation domination.

\begin{figure}[H]
  \centering
    \includegraphics[width=\linewidth]{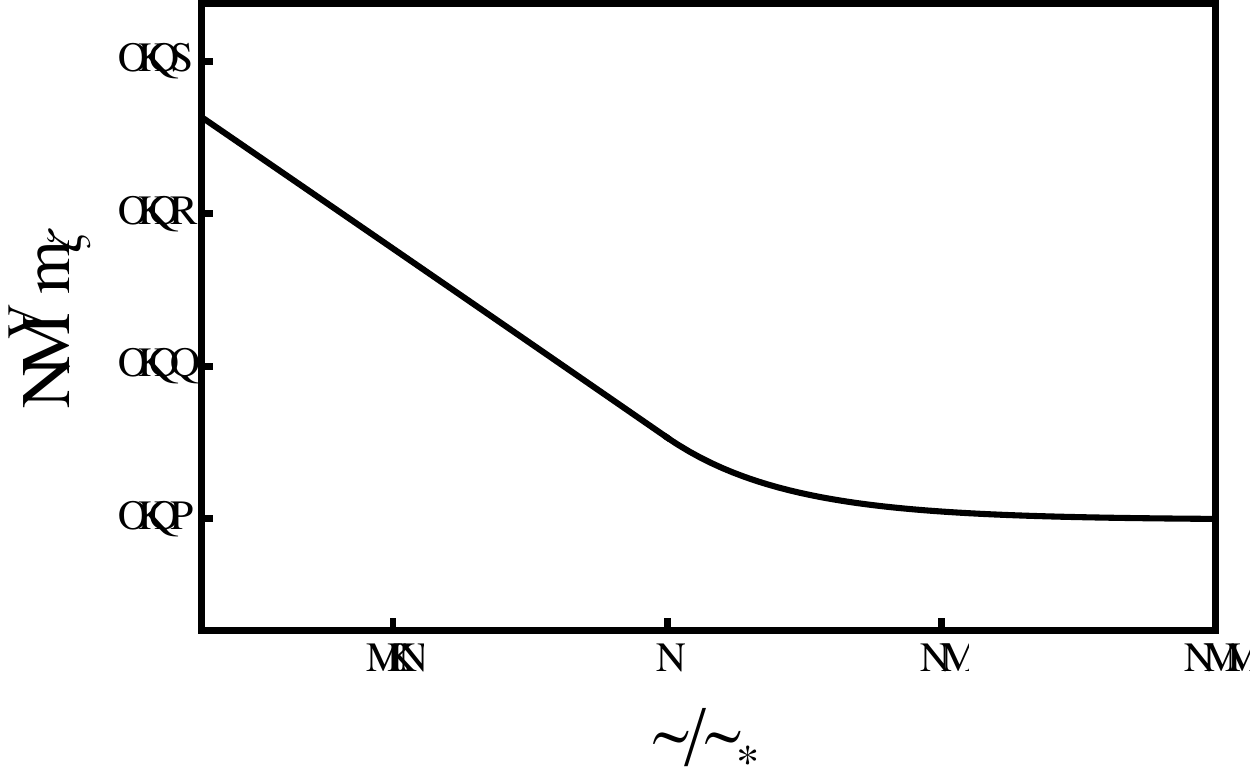}
	\caption{The power spectrum of $\zeta$ during the decay of the elastic solid to radiation for a superhorizon mode. During inflation, sound speed and equation of state parameters are chosen to be those listed in the third row of Table~\ref{exampleVals}.}
\label{P_zeta}
\end{figure}

To find the postinflationary scalar power spectrum, we match $\zeta$ and its first derivative at the time of the decay and will assume that the decay product is radiation. During radiation domination, on large scales $\zeta_{\textbf{k}}$ evolves as
\begin{equation}
\zeta_{\textbf{k}}'' + 2\mathcal{H}\zeta_{\textbf{k}}' \approx 0
\label{zetadecayeom}
\end{equation}
which has solution
\begin{equation}
\zeta_{\textbf{k}}(a\geq a_*) \approx \zeta_{\textbf{k}*} + \frac{\zeta_{\textbf{k}*}'}{\mathcal{H}_*}\left(1 - \frac{a_*}{a} \right)
\label{zetamatch}
\end{equation}
where integration constants have been chosen so that $\zeta_{\textbf{k}}$ and $\zeta_{\textbf{k}}'$ are continuous over the decay. From Eq.~(\ref{zetamatch}), we see that $\zeta_{\textbf{k}}$ has both a constant and decaying mode during radiation domination. Within a few $e$-folds after the decay, the decaying mode becomes negligible, as seen in Fig.~\ref{P_zeta}.

Using Eq.~(\ref{zetaR}) to calculate $\zeta$ from $R$ before the decay, for the case where $\epsilon_0=c_{s0}=0$, we find that the relationship between $\mathcal{P}_{\zeta}(k,a\gg a_*)$ and $\mathcal{P}_R(k,a_*)$ right before the decay, given in Eq.~(\ref{PRinf}), is
\begin{multline}
\frac{P_{\zeta}(k, a\gg a_*)}{\mathcal{P}_R(k, a\rightarrow ^-\!\!a_*)} = \bigg| \frac{3-4(3+2c_{s1}^2)\epsilon_1 + 4\tau_s + 2\tau_{\epsilon}}{3c_{s1}^4} \bigg|
\label{zetaRratio}
\end{multline}
As $c_{s1}<1$ (but not necessarily $c_{s1}\ll 1$), typically $\zeta_k$ will be larger than $R_k$, in which case if $\zeta_k$ is in the linear regime, so will $R_k$.

We can express the power spectrum of $\zeta$ in terms of the pivot scale $k_p=0.002 \, \text{Mpc}^{-1}$ as $P_{\zeta}(k)=A_{\zeta}(k/k_p)^{n_s-1}$. Assuming rapid reheating, for the $c_{s0}=\epsilon_0=0$ case $A_{\zeta}$ can be written as
\begin{equation}
A_{\zeta} = 10^{-22} \Delta \bigg| \frac{c_{s1}^{n_s-6}}{\epsilon_1} \bigg| \left(\frac{g_{RH}}{106.75}\right)^{\frac{7-n_s}{6}}  \left(\frac{T_{RH}}{10^{13} \, \text{GeV} }\right)^{5-n_s}
\label{Azeta}
\end{equation}
where $T_{RH}$ is the reheat temperature and $g_{RH}$ is the effective number of relativistic degrees of freedom contributing to the energy density at reheating.\footnote{In Eq.~(\ref{Azeta}),  we have assumed that all relativistic species are in thermodynamic equilibrium at $T_{RH}$ so that $g_{RH}=g_{sRH}$, where $g_s$ is the effective number of relativistic degrees of freedom contributing to the entropy density, and we make this assumption throughout the paper.} In the above expression, $\Delta$ is a constant that is given in detail in Appendix \ref{scalartensoramp}. For nearly scale-invariant scalar spectra, $\Delta$ is of order unity. When the sound speeds and equation of state are constant, $A_{\zeta}$ is given by the same expression except with the substitutions $c_{s1}\rightarrow c_s$ and $\epsilon_1\rightarrow \epsilon$.

Tracking the effect of a rapid decay to radiation on the tensor modes is straightforward, since as seen in Eq.~(\ref{heinst}) the tensor modes will only be affected by the change in anisotropic stress. Accordingly, we match the tensor perturbations and their first derivatives across the decay as the anisotropic stress vanishes. From Appendix~\ref{eomsection}, the equation of motion for the tensor perturbations is given by \footnote{In this section, $h^T$ will label a component of the tensor $h^T_{ij}$.}
\begin{equation}
(h_{\textbf{k}}^T)''+ 2\mathcal{H}(h_{\textbf{k}}^T)' + (k^2 + 4c_v^2\beta) h_{\textbf{k}}^T = 0
\label{hdecay}
\end{equation}
Since there is negligible rigidity in the radiation fluid, the transverse sound speed will vanish after the decay. During radiation domination, superhorizon tensor modes have the same approximate evolution equation as the scalar modes in Eq.~(\ref{zetadecayeom}); therefore, matching the tensor modes and their first derivatives across the decay will have the same form as the scalar mode solution in Eq.~(\ref{zetamatch}), so that
\begin{equation}
h_{\textbf{k}}^T (a\geq a_*) \approx h_{\textbf{k}*}^T + \frac{(h_{\textbf{k}*}^T)'}{\mathcal{H}_*}\left(1 - \frac{a_*}{a} \right)
\label{hmatch}
\end{equation}
Using the above equation with Eq.~(\ref{Phn}), the postinflationary tensor power spectrum of superhorizon modes is related to its value at the end of inflation by
\begin{equation}
	\frac{\mathcal{P}_T (k,a\gg a_*)}{\mathcal{P}_T (k, a_*)} = \frac{(1-2\gamma_T)^2 (\epsilon_0-1)^4}{4 (1-\epsilon_0+\epsilon_1)^2}
\end{equation}
where the tensor power spectrum at the end of inflation in given in Eq.~(\ref{PhH}).

The tensor-to-scalar ratio $r=\mathcal{P}_T/\mathcal{P}_{\zeta}$ after the decaying modes mentioned above are negligible for the $\epsilon_0=c_{s0}=0$ case can now be found to be
\begin{equation}
r \approx 16 c_{s1}^5 \epsilon_1 (a_*H_*/k)^{n_s - 1 - n_T}
\end{equation}
The tensor-to-scalar ratio is mildly dependent on the (physical) wavenumber and Hubble rate at the end of inflation, since in general the scalar and tensor modes have different tilts. At the pivot scale $k_p$, the tensor-to-scalar ratio is
\begin{multline}
r(k_p) \approx 1.94 \times 10^{22.9(n_s-0.96-n_T)} c_{s1}^5 \epsilon_1 \\\left( \left( \frac{g_{RH}}{106.75} \right)^{1/6}  \frac{T_{RH}} {10^{13} \, \text{GeV} }\right)^{n_s -1 - n_T}
\label{rapprox}
\end{multline}
For this case, the tensor-to-scalar ratio is suppressed by the $c_{s1}^5$ term, so for small values of $c_{s1}$ the tensor-to-scalar ratio will be highly suppressed. For example, with the parameter values in the first row of Table~\ref{exampleVals} with $c_{s1}=0.01$, the tensor-to-scalar ratio is $r \sim 10^{-12}$. However, if $c_{s1}$ assumes a higher value, then this suppression is more moderate, illustrated by the values in the second row of Table~\ref{exampleVals}, which with $c_{s1}=0.8$ yields a tensor-to-scalar ratio of $r=0.002$.

As in Section~\ref{slowroll}, we do not write out an explicit expression for the scalar amplitude or tensor-to-scalar ratio for cases when either $c_{s0}$ or $\epsilon_0$ are nonzero and will soon illustrate with numerical examples instead. But before doing this, we can gain some insight by examining the case when the sound speeds and equation of state are constant. As previously mentioned, a nearly scale-invariant scalar spectrum can be produced when $w$ is far from $-1$ if $c_s$ is sufficiently small. However, in this case there are added considerations as the energy density changes significantly during the course of inflation. If inflation lasts just long enough to solve the `horizon problem' (see Appendix \ref{efolds} for details) and $\Lambda$ is the energy scale at the beginning of inflation, $A_{\zeta}$ will be given by
\begin{multline}
A_{\zeta} \approx 10^{\frac{-109 - n_s(13.1+10.4\epsilon) + 30\epsilon}{2-\epsilon}} \Delta \bigg| \frac{c_s^{n_s-6}}{\epsilon} \bigg| \\ \times \left(\frac{g_{RH}}{106.75}\right)^{\frac{n_s-1+18\epsilon-4n_s\epsilon}{6(2-\epsilon)}} \left(\frac{\Lambda}{m_p}\right)^{\frac{5-n_s}{1+\epsilon/(2(1-\epsilon))}}
\end{multline}
for $c_s$, $\epsilon$ constant, where $m_p$ is the Planck mass. For solutions with $n_s\sim1$, if $\epsilon$ is raised to higher values, $A_{\zeta}$ may drop significantly. If $\Lambda$ is  bounded by the Planck scale, for large values of $\epsilon$ ($w$ far from $-1$), to keep $A_{\zeta}\sim 10^{-9}$, $c_s$ may have to be fine-tuned to a very small value. Alternatively, this fine-tuning may be averted if one is comfortable having inflation start at supra-Planckian scales.

This issue is demonstrated for the slow-varying sound speeds case in the fifth row in Table~\ref{exampleVals}, in which $w$ varies close to $-2/3$. To attain the same scalar amplitude as was used in the other examples in Table~\ref{exampleVals} and have inflation start at or below the Planck scale and last for a sufficiently long duration, $c_s$ had to assume the extremely small value of $\sim 10^{-8}$. In the fourth row of Table~\ref{exampleVals}, $w \simeq -5/6$, so while the departure from $-1$ is still significant, we find $c_s$ can assume much larger values near $10^{-3}$ --- and smaller values of $w$ may have correspondingly larger values of $c_s$.

However, even if we do not tune $c_s$ to be very small, it is theoretically interesting that we can achieve a scale-invariant spectrum far from $w \simeq -1$ even if the amplitude of perturbations are not large enough to match observations.  For instance, the parameter values $w=-2/3$ and $c_s=1/10$ produce a slightly blue-tilted spectrum for both scalar and tensor perturbations (with $n_s\approx1.04$ and $n_T\approx 0.04$).  Understanding the physical origin of this near scale invariance is an extremely interesting question that might give new insight into the physics of horizons.

Lastly, we compute the value of the tensor-to-scalar ratio for our examples where either $c_{s0}$ or $\epsilon_0$ are nonzero. Using the values in the fourth row of Table~\ref{exampleVals}, where $w$ varies slowly near $-5/6$, gives a tensor-to-scalar ratio at the pivot scale of $r\sim 10^{-15}$. We again see that the tensor-to-scalar ratio is highly suppressed by $c_{s1}$. With the values listed in the last row of Table~\ref{exampleVals} with $c_{s0}=0.15$ yields $r\approx 10^{-6}$ and the suppression of $r$ is more moderate.

\section{Conclusion}

By having a sufficiently rigid structure, a relativistic elastic solid is capable of driving an inflationary stage in the early Universe. In the case of constant sound speeds $c_s$ and $c_v$ and equation of state $w$, a blue-tilted scalar power spectrum is produced. Allowing the sound speeds and equation of state to vary slowly in time can result in a red-tilted scalar power spectrum with small running. When $c_s$ is small, the tensor-to-scalar ratio will be highly suppressed, but can attain larger values for higher values of $c_s$.

An interesting feature of this model is that perturbations evolve on superhorizon scales, even in the absence of nonadiabatic pressure. The superhorizon evolution results from the shear stresses in the solid, where the propagation of a single perturbative mode causes an anisotropic pressure. Because of this anisotropy, when smoothed on a superhorizon scale, different locations in the Universe will not share the same FRW evolution, as they do when both shearing stresses and nonadiabatic pressures are absent. As a result, the perturbations do not `freeze-out' soon after horizon crossing and consequently, the details of the end of inflation can impact both scalar and tensor power spectra for modes that are on superhorizon scales when inflation ends. The case of a rapid decay of the elastic solid into radiation was explored as a specific example.

Finally and intriguingly, we find this model allows for $w$ to vary slowly near values that are significantly different from $-1$ and can find cases where this produces nearly scale-independent scalar and tensor power spectra despite being far from the de Sitter regime.  This is surprising and unexpected and it would be interesting to determine the underlying physical reason for this phenomena.

\acknowledgments

This research was supported by the National Science and Engineering Research Council of Canada (NSERC). KS is supported in part by a NSERC of Canada Discovery Grant. 
While this work was in its final stages, another closely related paper, Ref.~\cite{solidInflation}, appeared.   While Ref.~\cite{solidInflation}  studies essentially the same system as we do here it uses an entirely different formalism than we do (an effective field theory for the elastic solid), makes interesting predictions for non-Gaussianity, and amongst other things makes important contrasts on how this system is distinct from the EFT of inflation.   Despite the fact that we take completely independent approaches to the problem we have verified that our results agree in the $w \simeq -1$ limit that Ref.~\cite{solidInflation}  emphasizes.  Our approach directly quantizes the quadratic action for the elastic solid system, has an extended treatment of reheating and superhorizon evolution, and has been explicitly applied in both the $w \simeq -1$ and far from $w \simeq -1$ regimes.  After the completion of our paper another work, Ref.~\cite{anisotropicSolid}, appeared nearly simultaneously to ours that, in the language of Ref.~\cite{solidInflation}, discusses superhorizon evolution and anisotropy in an inflating elastic solid with similar results to ours for the evolution of linear modes.  We would like to thank Adam Brown, Daniel Green, Matthew Kleban, Andrew Pontzen, and Matias Zaldarriaga for their comments, discussions, and feedback.

\appendix

\section{Equations of Motion for Scalar and Tensor Perturbations}
\label{eomsection}

In Section~\ref{actionSection}, we derived the equations of motion for the mode functions $\chi$ and $X$ for the scalar and tensor linear perturbations, respectively, from the action, which can then be used to find the equations of motion for $R$ and $h^T_{ij}$. Alternatively, we can derive these equations of motion directly from the Einstein equations and the properties of the elastic solid given in Eq.~(\ref{elasticParams}). Although we can derive the equations of motion for $u$ and $U_p$ in this manner, it is only through the action in which we can properly identify $u$ and $U_p$ as the canonical variables for the scalar and tensor linear perturbations, respectively.

As before, to derive the equations of motion for the scalar perturbations, it is convenient to work in the comoving gauge and will also move to Fourier space. We can combine the Fourier space versions of Eqs.~(\ref{einst00co}) and (\ref{veqn2}) to eliminate $\delta$ and then combine the derivative of this equation with Eqs.~(\ref{einst00}), (\ref{einstii}), and (\ref{einstij}) in the comoving gauge to eliminate $\delta$, $E'$ and $E''$, then use Eq.~(\ref{einst0ico}) and its derivative to eliminate $\phi$ and $\phi'$, which gives
\begin{multline}
\psi_{\textbf{k}}'' + \left[ \left(2+3\left(w-\frac{d P}{d \rho}\right)\right)\mathcal{H} -\left(\text{ln}\left(\frac{d P}{d \rho}\right)\right)' \right]\psi_{\textbf{k}}' \\ + \frac{d P}{d \rho}k^2\psi_{\textbf{k}} +  \frac{\mathcal{H}}{3(1+w)\frac{d P}{d \rho}}\left[ 3\mathcal{H}\frac{d P}{d \rho}\left(3w+w^2-2\frac{d P}{d \rho}\right) \right. \\ \left. -2w\left(\frac{d P}{d \rho}\right)' \right]\Pi_{\textbf{k}} + \frac{2}{3}\frac{w}{1+w}\mathcal{H}\Pi_{\textbf{k}}' = 0
\label{psipieqn}
\end{multline}
This result does not assume any properties of the substance occupying our spacetime, other than being able to write its stress-energy tensor in the form in Eq.~(\ref{Tparam}).

We now specify the elastic properties of our substance by employing Eq.~(\ref{pieqn2}), which can be written in terms of $\psi_{\textbf{k}}$ by use of the Fourier version of Eq.~(\ref{Piscalar}). Substituting this into Eq.~(\ref{psipieqn}) yields
\begin{equation}
R_{\textbf{k}}'' + 2\frac{z'}{z}R_{\textbf{k}}' + \left[ c_s^2k^2 + m^2_{\text{eff},S} + \frac{z''}{z} \right]R_{\textbf{k}} = 0
\end{equation}
where $z$ and $m^2_{\text{eff},S}$ were defined in Eqs.~(\ref{zdef}) and (\ref{meff}), respectively, and have used the fact that $R=\psi$ in the comoving gauge.  By substituting $u$ for $R$ using Eq.~(\ref{udef}), we can recover the equation of motion for $u$ in Eq.~(\ref{ueom}).

Obtaining the equation of motion for the tensor perturbations is very straightforward, since there is only one Einstein equation for the tensor perturbations, given in Eq.~(\ref{heinst}). With the tensor part of the anisotropic stress for the elastic solid in Eq.~(\ref{Pitensor}), the equation of motion for the tensor perturbations is
\begin{equation}
(h^T_{\textbf{k}})^{i\prime\prime}_j + 2\mathcal{H}(h^T_{\textbf{k}})^{i\prime}_j + (k^2 + 4c_v^2\beta) (h^T_{\textbf{k}})^i_j = 0
\end{equation}
where we have switched to Fourier space.

\section{The `Horizon Problem' Revisited}
\label{efolds}

An interesting feature of this model of inflation is that it allows  the possibility of far from de Sitter backgrounds that nevertheless produce a nearly scale-invariant two-point correlation function. In such a case, the horizon may change significantly during inflation, thus altering the `horizon problem', in which we expect to be able to fit the present-day horizon size into the horizon at the beginning of inflation expanded to today. Labelling quantities evaluated at the beginning of inflation, reheating, and the present-day by the subscripts $i$, ${RH}$, and $0$, respectively, we require
\begin{align}
H_0^{-1} &\leq \frac{a_0}{a_i}  H_i^{-1} \nonumber \\
&= \frac{a_0}{a_{RH}}\frac{a_{RH}}{a_i}   H_i^{-1} \nonumber \\
&\approx \left(\frac{g_{sRH}}{g_{s0}}\right)^{1/3} \frac{T_{RH}}{T_0} e^N   H_i^{-1}
\end{align}
where $N$ is the number of $e$-folds of inflation and $g_s$ is the effective number of relativistic degrees of freedom contributing to the entropy density.

If the equation of state during inflation is $w$, then assuming $w$ is approximately constant, the horizon size changes as $\propto a^{3(1+w)/2}$ during inflation, so that $H_i^{-1} \approx e^{-3(1+w)N/2} H_{RH}^{-1}$. The minimum number of $e$-folds of inflation to solve the `horizon problem' then becomes
\begin{align}
N &\geq \frac{1}{1-\epsilon} \left( \text{ln}\left(\frac{T_0}{H_0}\right) + \text{ln}\left(\frac{H_{RH}}{T_{RH}}\right) + \frac{1}{3}\text{ln}\left(\frac{g_{s0}}{g_{sRH}}\right) \right) \nonumber \\
&\approx \frac{1}{1-\epsilon} \left( 54 + \text{ln}\left(\frac{T_{RH}}{10^{13} \, \text{GeV}}\right) + \frac{1}{3}\text{ln}\left(\frac{106.75}{g_{sRH}} \right)  \right)
\label{NminHP}
\end{align}
where $\epsilon=3(1+w)/2$ and we have taken $g_{s0}=43/11$. Thus, the minimum number of $e$-folds required when $w$ is far from $-1$ ($\epsilon$ is large) may be substantially larger than that for the standard $w\simeq -1$ case.

If $\Lambda$ is a high-energy cut-off scale that is an upper bound for the initial energy density of inflation (presumably the Planck scale), then for $\epsilon\neq 0$, $N$ is bounded from above by
\begin{multline}
N \lesssim \frac{1}{2\epsilon} \left(172 + 4\left(\text{ln}\left(\frac{\Lambda}{m_p}\right) - \text{ln}\left(\frac{T_{RH}}{\text{GeV}}\right) \right) \right. \\ \left. - \text{ln}\left(\frac{g_{RH}}{106.75}\right) \right)
\end{multline}
where $m_p=\sqrt{8\pi/3} \, l^{-1}$ is the Planck mass and $g_{RH}$ is the effective number of relativistic degrees of freedom contributing to the energy density.  If the bound from Eq.~(\ref{NminHP}) provides the strongest constraint on the minimum number of $e$-folds of inflation, which will likely be the case if $w$ is far from $-1$, then the maximum reheat temperature such that $N$ is appropriately bounded is
\begin{multline}
\text{log}_{10}\left(\frac{T_{RH}}{\text{GeV}}\right) \approx \frac{1}{1-\epsilon/2}\bigg( 19 - 24\epsilon  \\ \left. + 0.44(1-\epsilon)\text{ln}\left(\frac{\Lambda}{m_p}\right) + (0.18\epsilon - 0.11)\text{ln}\left(\frac{g_{RH}}{106.75}\right) \right)
\end{multline}
where we have assumed $g_{RH}=g_{sRH}$.

\section{Multicomponent System with Energy-Momentum Transfer}
\label{decay}

In this appendix, we review the equations governing the energy-momentum transfer between multiple `fluid-like' substances.\footnote{By `fluid-like', we are not referring to a perfect fluid, but instead any substance whose stress-energy tensor can be parametrized by Eq.~(\ref{Tparam}), which includes an elastic solid.} This discussion will follow the work of Ref.~\cite{interactingFluids}. We will begin by stating the general equations for energy-momentum transfer between any number of `fluid-like' substances and then specialize to the case of an elastic solid decaying into a perfect fluid. As in Section~\ref{reheating}, we will use the coordinate time instead of conformal time in this section.

The energy-momentum transfer 4-vector $Q_{(\alpha)}^{\nu}$ that appears in Eq.~(\ref{Talpha}) must vanish when summed over all substances $\alpha$
\begin{equation}
\sum_{\alpha} Q_{(\alpha)}^{\nu} = 0
\label{Qsum}
\end{equation}
for the total stress-energy tensor to be covariantly conserved. The conservation equation for the background energy density $\rho_{\alpha}$ of substance $\alpha$ is
\begin{equation}
\dot{\rho}_{\alpha} = -3H(\rho_{\alpha}+P_{\alpha}) + Q_{\alpha}
\end{equation}
where $Q_{\alpha}=Q_{(\alpha)}^0$. Importantly, from Eq.~(\ref{Qsum}), the conservation equation for the total energy density given in Eq.~(\ref{drho}) still holds.

If all substances have vanishing intrinsic nonadiabatic pressure ($\delta P_{\alpha} = (d P_{\alpha}/d\rho_{\alpha}) \delta\rho_{\alpha}$) then the equations of motion for $\zeta_{\alpha}$ and $R_{\alpha}$, defined in Eqs.~(\ref{zetaalpha}) and (\ref{Ralpha}), are given by
\begin{multline}
\dot{\zeta}_{\alpha} = -\frac{H}{\dot{\rho}_{\alpha}}(\delta Q_{\text{intr},\alpha} +\delta Q_{\text{rel},\alpha} ) - \frac{\dot{H}}{H}(R-\zeta) \\ + \frac{k^2}{3a^2H} \left( 1 - \frac{Q_{\alpha}}{\dot{\rho}_{\alpha}} \right) R_{\alpha}
\end{multline}
and
\begin{multline}
\dot{R}_{\alpha} = \frac{\dot{H}}{H}(R_{\alpha}-R) - \frac{\dot{\rho}_{\alpha}}{\rho_{\alpha}+P_{\alpha}} \frac{d P_{\alpha}}{d \rho_{\alpha}}(R_{\alpha}-\zeta_{\alpha}) \\ - \frac{H}{\rho_{\alpha}+P_{\alpha}}\left(\frac{2}{3}P_{\alpha}\Pi_{\alpha} + f_{\text{rel},\alpha} \right)
\end{multline}
where we have written the equations in Fourier space, but neglected to write the subscript $k$ to minimize the number of subscripts written on each variable. In the above equations, $\delta Q_{\text{intr},\alpha}$ and $\delta Q_{\text{rel},\alpha}$ are the intrinsic and relative nonadiabatic energy transfer perturbations, respectively, and $ f_{\text{rel},\alpha}$ is the relative momentum transfer, which are defined as
\begin{subequations}\label{grp}
	\begin{align}
		\delta Q_{\text{intr},\alpha} &\equiv \delta Q_{\alpha} - \frac{\dot{Q}_{\alpha}}{\dot{\rho}_{\alpha}}\delta\rho_{\alpha} \\
		\delta Q_{\text{rel},\alpha} &\equiv - \frac{Q_{\alpha}}{6H\rho}\sum_{\beta} \dot{\rho}_{\beta} \mathcal{S}_{\alpha\beta} \\
		f_{\text{rel},\alpha} &\equiv a Q_{\alpha} \sum_{\beta} \frac{\rho_{\beta}+P_{\beta}}{\rho + P} v_{\alpha\beta}
	\end{align}
\end{subequations}
where $ \delta Q_{\alpha}$ is the perturbation to $Q_{\alpha}$.

With $Q_e^{\nu}$ for the decay of the elastic solid specified in Eq.~(\ref{Qdecay}), we have $Q_e = -Q_f = -\Gamma \rho_e(1+w_e)$. Since $Q_e$ is a function of $\rho_e$, $\delta Q_{\text{intr},e} =0$. Using this fact and $\delta Q_f = -\delta Q_e$, we find that $\delta Q_{\text{intr},f} =\dot{Q}_e \mathcal{S}_{ef}/3H$.

\vspace{3mm}

\section{Scalar Amplitude}
\label{scalartensoramp}

In this appendix we give detailed expressions for the scalar amplitude evaluated at the pivot scale $k_p$, as given in Eq.~(\ref{Azeta}). 
We first examine the case where the sound speeds and equation of state are varying slowly in time and  $\epsilon_0=c_{s0}=0$. From Eqs.~(\ref{PRinf}) and (\ref{zetaRratio}), $\Delta$ is given by
\begin{multline}
 \Delta =  0.11 \times 10^{5.23(n_s-0.96) }  \Gamma^2(\gamma_S) \\ (3-18\epsilon_1 - 8c_{s1}^2\epsilon_1-2\tau_s + 2\tau_{\epsilon}) \\ \times g_{s0}^{\frac{1-n_s}{3}} (T_0/k_p)^{1-n_s}
\end{multline}
where we have assumed that the reheating process is very rapid. Choosing the pivot scale as $k_p=0.002 \, \text{Mpc}^{-1}$, and using the CMB temperature $T_0=2.725 \, K$ and $g_{s0}=43/11$, $\Delta$ becomes
\begin{multline}
 \Delta =  1.53 \times 10^{- 28.5(n_s-0.96) } \Gamma^2(\gamma_S) \\  (3-18\epsilon_1 - 8c_{s1}^2\epsilon_1-2\tau_s + 2\tau_{\epsilon})
\end{multline}

In the case of constant sound speeds and equation of state, $\Delta$ is found to be
\begin{multline}
 \Delta =  5.82\times 10^{-5 - 22.9(n_s-1)} \Gamma^2(\nu) \\ (21 + n_s(n_s-10) )^2 |1+3w|^{7-n_s}
\end{multline}


\begin{thebibliography}{9}

\bibitem{Bucher}
  M.~Bucher and D.~N.~Spergel,
  Phys.\ Rev.\  D {\bf 60}, 043505 (1999)
  [arXiv:astro-ph/9812022].

\bibitem{MossPert}
  R.A.~Battye and A.~Moss,
  Phys.\ Rev.\  D {\bf 76} 023005 (2007).

\bibitem{massiveGravity}
R.A.~Battye and A. J. Pearson,
arXiv:1301.5042.

\bibitem{Gruz}
  A.~Gruzinov,
  Phys.\ Rev.\  D {\bf 70}, 063518 (2004).

\bibitem{solidInflation}
S.~ Endlich, A.~Nicolis and J.~ Wang,
arXiv:hep-th/1210.0569.

\bibitem{anisotropicSolid}
N.~Bartolo, S.~Matarrese, M.~Peloso and A.~Ricciardone,
arXiv:1306.4160.

\bibitem{CarterFoundations}
  B.~Carter and H.~Quintana,
  Proc.\ R.\ Soc.\ A {\bf 331} 57 (1972).

\bibitem{rigidity}
  R.~A.~Battye, B.~Carter, E.~Chachoua and A.~Moss,
  Phys.\ Rev.\  D {\bf 72}, 023503 (2005)
  [arXiv:hep-th/0501244].

\bibitem{domainwalls}
  R.~A.~Battye, E.~Chachoua and A.~Moss,
  Phys.\ Rev.\  D {\bf 73}, 123528 (2006)
  [arXiv:hep-th/0512207].

\bibitem{Landau}
L.~D.~Landau and E.~M.~Lifshitz,
{\it Theory of Elasticity}, (Pergamon Press Ltd., New York, 1986), 3rd ed.

\bibitem{CarterSpeed}
B.~Carter,
Phys.\ Rev.\  D {\bf 7}, 1590 (1973).

\bibitem{Fock}
V.~Fock,
{\it The Theory of Space, Time and Gravitation}, (Pergamon Press Ltd., Oxford, 1964), 2nd rev. ed.

\bibitem{Mukhanov}
V.F.~Mukhanov, H.A.~Feldman and R.H.~Brandenberger,
Phys. Rept. {\bf 215}, 203 (1992).

\bibitem{separateUniverse}
D.~Wands, K.~A.~Malik, D.~H.~ Lyth and A.~R.~Liddle,
Phys. Rev. D {\bf 62}, 043527 (2000).

\bibitem{conservedCosmologicalPerturbations}
D.~H.~ Lyth and D.~Wands, 
Phys. Rev. D {\bf 68}, 103515 (2003).  

\bibitem{cosmologicalModels}
G.~F.~R. ~Ellis and H.~van Elst,
``Cosmological Models",
{\it Theoretical and Observational Cosmology: Proceedings of the NATO Advanced Study Institute on Theoretical and Observational Cosmology}.
NATO Science Series C. {\bf 541} (Kluwer Academic, Boston, 1999), p.1-116

\bibitem{bianchi}
A.~ Pontzen and A.~Challinor, 
Class. Quant. Grav. {\bf 28}, 185007 (2011).

\bibitem{wmap9}
G.~ Hinshaw et al.,
arXiv:astro-ph/1212.5226.

\bibitem{interactingFluids}
 K.~A.~Malik and D.~Wands,
 JCAP {\bf 2} 7 (2005).

\bibitem{baumann}
D.~Baumann, L.~Senatore, and M.~Zaldarriaga,
JCAP {\bf 5}, 4 (2011).

\bibitem{weinberg}
S.~ Weinberg, 
Phys. Rev. D {\bf 67}, 123504 (2003).  

\end{thebibliography}
\end{document}